# Joint routing and pricing control in congested mixed autonomy networks

Mohammad Hadi Mansourianfar[a], Ziyuan Gu[b], S. Travis Waller[a], Meead Saberi[a,*]

[a] *Research Center for Integrated Transport Innovation (rCITI), School of Civil and Environmental Engineering, University of New South Wales (UNSW) Sydney, NSW 2052, Australia*
[b] *Jiangsu Key Laboratory of Urban ITS, Jiangsu Province Collaborative Innovation Center of Modern Urban Traffic Technologies, School of Transportation, Southeast University, Nanjing 210096, China*

**Abstract**
Routing controllability of connected and autonomous vehicles (CAVs) has been shown to reduce the adverse effects of selfish routing on the network efficiency. However, the assumption that CAV owners would readily allow themselves to be controlled externally by a central agency for the good of the system is unrealistic. In this paper, we propose a joint routing and pricing control scheme that aims to incentivize CAVs to seek centrally controlled system-optimal (SO) routing by saving on tolls while user equilibrium (UE) seeking human-driven vehicles (HVs) are subject to a congestion charge. The problem is formulated as a bi-level optimization program where the upper level optimizes the dynamic toll rates using the network fundamental diagram (NFD) and the lower level is a mixed equilibrium simulation-based dynamic traffic assignment model (SBDTA) considering different combinations of SO-seeking CAVs. We apply a feedback-based controller to solve for the optimal spatially differentiated distance-based congestion charge from which SO-seeking CAVs are exempt; but UE-seeking HVs are subject to the charge for entering the city center. To capture the distinct microscopic behavior of CAVs in the mixed autonomy traffic, we also implement an adaptive link fundamental diagram (FD) within the SBDTA model. The proposed joint control scheme encourages CAV owners to seek SO routing resulting in less total system travel time. It also discourages UE-seeking HVs from congesting the city center. We demonstrate the performance of the proposed scheme in both a small network and a large-scale network of Melbourne, Australia.

**Keywords:** Congestion pricing; Simulation-based dynamic traffic assignment; Mixed equilibrium traffic assignment; Autonomous vehicles; Bi-level optimization.

## 1. Introduction

The advent of connected and autonomous vehicles (CAVs) is expected to change people's travel behavior and network traffic flow characteristics in many aspects (Fagnant and Kockelman, 2015). Previous studies have shown that CAVs could enhance network flow capacity (Levin and Boyles, 2016), improve traffic operations at freeway mainlines (Han et al., 2021), ramps (Dong et al., 2019) or signalized intersections (Le Vine et al., 2016) among others. On the other hand, reduced reaction times, closer spacing between vehicles, and increased demand for making trips by private cars will likely exacerbate the traffic congestion in some parts of the network (Simoni et al., 2019, Levin and Boyles, 2016). So positive and negative aspects of this new technology should be studied and adequately addressed.

Before human-driven vehicles (HVs) are completely phased out, the urban traffic flow will still be heterogeneous of both human-driven and self-driving vehicles, commonly known as mixed autonomy. The potential of CAVs in managing urban road congestion is substantial not only because of vehicle-specific attributes such as less safe time headway realized by marginal reaction time from computers (Friedrich, 2016), but also due to the connectivity or communication capabilities that make these vehicles congestion-aware and controllable (Rossi et al., 2017). As such, system optimum (SO)-seeking CAVs are conceivable as opposed to user equilibrium (UE)-seeking selfish HVs.

One critical component of deploying SO-seeking CAVs is a traffic management center that monitors and fully controls these vehicles in real time (Zhang and Nie, 2018). However, it is questionable that CAV owners or operators would allow themselves to be controlled externally by a central agency for the good of the system. In this research, we aim to utilize a congestion pricing scheme as an economic lever (or incentive for central control), from which SO-seeking CAVs are exempt while UE-seeking HVs have their usual shortest-path routing decisions but are subject to a spatially differentiated congestion charge. This control strategy could potentially boost market penetration rate of CAVs and then encourage them to adopt SO routing behavior (brings less total system travel time (TSTT)) and discourage UE-seeking users from entering the congested areas like central business district (CBD) of the cities. Previously studied ideas such as allocating CAV only zones (Chen et al., 2017), designing automated driving sub-networks (Madadi et al., 2020) or AVs expressways (Wu et al., 2020) may not be practical in real-world as they may affect HVs' accessibility to destinations. Unlike previous studies on congestion pricing of CAVs (Sharon et al., 2017, Simoni et al., 2019), this research aims to consider and model both SO-seeking CAVs and UE-seeking HVs (i.e. different routing strategies) in a dynamic and second-best priced traffic network.

The objective of this research is twofold. We first aim to develop a mixed equilibrium simulation-based dynamic traffic assignment (SBDTA) framework for a mixed fleet of CAVs and HVs based on seminal studies of (Mahmassani and Peeta, 1993) and (Mahmassani, 1994), where UE-seeking HVs follow the standard shortest travel time paths whereas SO-seeking CAVs follow the shortest marginal travel time paths identified by the central agency. Building upon the mixed equilibrium framework, we then aim to develop a feedback-based spatially differentiated congestion pricing controller utilizing the macroscopic or network fundamental diagram (NFD or MFD) (Geroliminis and Daganzo, 2008), inspired by Zheng et al. (2016) and Gu et al. (2018b), which applies to UE-seeking HVs. The resulting joint routing and pricing control serves as a hybrid network traffic optimizer that could potentially bring significant benefits to the whole system and the CBD. To the best of the authors' knowledge, such a mixed equilibrium joint control framework of CAVs has never been proposed before and hence offers significant contributions to the literature.

The remainder of this paper is structured as follows. Section 2 presents a review of relevant literature. The proposed and developed methodologies are presented in Section 3 including a simulation-based solution algorithm for the mixed equilibrium dynamic traffic assignment (DTA) problem, the NFD-based PI feedback pricing control strategy, and the mathematical formulation of joint routing and pricing control. In Section 4, we present the numerical experiments to demonstrate the performance of the proposed framework and its scalability to a large-scale network. Finally, section 5 presents conclusions and a few directions for future research.

## 2. Literature review

### 2.1. Multiclass traffic assignment models

CAVs are expected to operate on the roads alongside HVs over the next decade or so (Litman, 2017). Therefore, different route choice behaviors will likely be perceived in the network. According to Wardrop's principles, equilibrium in traffic assignment can be achieved through two route choice strategies: UE and SO (Wardrop and Whitehead, 1952). The third behavior frequently adopted to capture HVs behavior is stochastic user equilibrium (SUE). Having different knowledge levels of prevailing traffic conditions introduces stochasticity and uncertainty in making route choices. UE and SUE behavior is prevailing among HVs, while SO strategy can be followed by a specific class of vehicles like CAVs that are centrally controlled by a traffic management center (TMC) to reduce selfish and inefficient behavior (Yang, 1998). Previous studies considered different route choice behaviors for equipped and non-equipped vehicles including SO, deterministic or stochastic UE (Harker, 1988, Van Vuren and Watling, 1991, Bennett, 1993, Ben-Akiva et al., 1991, Yang, 1998). Harker (1988) introduced the mixed equilibrium traffic assignment for the first time and considered two classes including SO and UE users and employed variational inequality to solve the problem for a small network. Van Vuren and Watling (1991) considered stochasticity, whereas unequipped vehicles with advanced traveler information system (ATIS) have higher error rate compared to guided travelers who do not follow the directions completely. Huang and Li (2007) presented a multicriteria (cost and time) logit-based SUE assignment model for multiclass traffic flow. Users of each class, differentiated by a specific value of time (VOT), were divided into two subclasses, equipped and unequipped with ATIS. The equipped users were considered to have lower perception error for estimating the travel cost compared to unequipped vehicles. The model was formulated as a fixed-point problem and solved by the widely used method of successive averages (MSA) algorithm with a predefined step size converging to zero (Sheffi, 1985). Their results showed that overreaction occurs more easily on higher VOT user classes. Overreaction occurs when a substantial fraction of drivers receives and responds to descriptive information on traffic conditions that may lead to congestion due to shifting from their own routes to the recommended one (Ben-Akiva et al., 1991, Huang and Li, 2007).

Recently mixed equilibrium traffic assignment models have been adopted to model CAVs and HVs such that to reflect the benefits of self-driving vehicles in a network. Intelligent vehicles have the potential to be controlled by some central authority that assigns the routes. A number of previous studies have modeled HVs and CAVs to follow SUE, UE, or SO route choice strategies (Wang et al., 2019, Zhang and Nie, 2018, Bagloee et al., 2017). Wang et al. (2019) proposed a static multiclass traffic assignment model in which SUE-seeking HVs and UE-seeking CAVs are characterized by a cross-nested logit (CNL) model and a user equilibrium model, respectively. The CNL model is used to reflect the HVs' uncertainty associated with limited knowledge of traffic conditions overcoming the flow overestimation among routes with common links (i.e. overlapping routes issue of the logit-based SUE). To capture the benefits of platooning and lower reaction times of CAVs, Levin and Boyles (2015) proposed a multiclass, four-step macroscopic model that considers increase in link capacity as a function of the proportion of CAVs on the links. However, they used UE traffic assignment approach for both CAVs and HVs. A novel model was proposed by Chen et al. (2017) to deal with mixed-routing behaviors in the network. They developed a mathematical framework to optimally identify zones where only CAVs can enter and must apply SO routing principle within the CAV zones. Like HVs, CAVs can follow UE strategy outside of the designated zones. The main shortcoming of these modeling approaches is that their investigations were limited to the static multiclass traffic assignment models that generally depict the mixed traffic flow in a steady state, thus is unable to reflect the system and drivers' time-dependent behaviors, the queue charge and discharge, congestion propagation and dissipation, and other traffic flow phenomena that happen over time (Daganzo, 1998).

Two approaches exist in academic literature to deal with the dynamic traffic assignment (DTA) problem, analytical and simulation-based, in which origin-destination trips are split into different departure time intervals to provide time-dependent estimates of traffic flow patterns (Chiu et al., 2011, Wang et al., 2018). The analytical models replicate the various dynamic traffic flow phenomena by adding explicit constraints in the DTA formulations over the time and space realm that is a complex task especially in large-scale networks. In a seminal work conducted by (Mahmassani and Peeta, 1993) and (Peeta and Mahmassani, 1995), a simulation-based solution algorithm was proposed to circumvent the need for link cost and link exit functions to solve the DTA problem. The simulation-based DTA models have been growing particularly for modeling and analyzing large-scale transportation systems for operational and planning purposes (Mahmassani, 2001, Ziliaskopoulos et al., 2004, Shafiei et al., 2018).

In Mahmassani and Peeta (1993), an iterative dynamic simulation-based solution algorithm was only applied to two scenarios (100% UE- and 100% SO-seeking conventional vehicles with the same vehicle-specific attributes) in which MSA was used for finding the descent direction in the search process in which a simple convergence criterion was used: a path was considered converged if the relative change in the number of vehicles assigned to it at the current and previous iterations becomes relatively small; if the proportion of converged paths across all the OD and assignment intervals reaches an acceptable threshold, the equilibrium state is considered being achieved. Hu et al. (2018) also developed a simulation-based DTA framework to solve mixed traffic flow problem considering four vehicle types (car, bus, motorcycle, and truck) and four different assignment rules (pre-specified-path, UE, SO, and real-time information). The convergence speed of the solution algorithms relying on MSA is generally slow when the iteration number is large, because the generated auxiliary flow slightly contributes to find the solution point. This drawback is mainly due to the predefined step size sequence converging to zero, which has inspired researchers to develop alternative search direction methods. Frank and Wolf (FW) is an optimized step size algorithm that provides better convergence speed than MSA. The advantage of the FW and its variants in solving the traffic assignment problem is that there is no need to store paths information, which is a significant benefit in terms of memory requirements especially for large-scale networks. However, the performance function of the links in the network must be known (Frank and Wolfe, 1956); therefore, it is not readily applicable in simulation based DTA problems. Method of successive weighted average (MSWA) is another option that has the simplicity of MSA and can be applied in simulation based DTA problems. MSWA determines a new step size that gives higher weight to the auxiliary flow from the previous iterations (Liu et al., 2009).

### 2.2. Congestion pricing models

Besides all the potential benefits of the CAV technology in improving traffic flow, increased travel demand by private cars will likely exacerbate the traffic congestion (Simoni et al., 2019, Levin and Boyles, 2016, Fagnant and Kockelman, 2015, Wadud et al., 2016). Congestion pricing, as one of the most effective travel demand management (TDM) policies, can mitigate the issue. Advances in CAV and connectivity technologies offer a unique opportunity to implement more complex, efficient and behaviorally effective congestion pricing strategies that vary over time and space (Gu et al., 2018a, Simoni et al., 2019).

Two widely used congestion pricing models in the literature include first-best and second-best models. In the first-best or Pigouvian model, travelers must pay a toll equivalent to the marginal external cost to make the generalized cost of a trip equal to the marginal social cost (Pigou, 1920, Verhoef, Yang and Huang, 2005). Despite solid theoretical basis, imposing toll to all links results in huge operational costs and public acceptance concerns; therefore, it has only been applied to small networks for demonstration purposes (Yang and Huang, 2004). On the other hand, a variety of second-best pricing schemes have been introduced in which only a sub-area of the network is tolled. Several studies in the literature have already provided frameworks to find optimal toll locations and toll levels (Fan, 2016, Sun et al., 2016, Aboudina and Abdulhai, 2017). From the modeling perspective, the second-best pricing problem has been modeled widely as a bi-level optimization problem or equivalently mathematical programming with equilibrium constraints (MPEC) (Liu et al., 2013, Liu et al., 2014, Yang and Zhang, 2003,

Zhang and Yang, 2004). The upper level aims to minimize the total system travel time or maximizes the total social welfare by considering a specific toll level and the lower level is usually a conventional static UE or SUE traffic assignment problem with elastic or fixed demand. However, Solving the MPEC of DTA (with multiple time-dependent OD matrices) and for a large-scale network is still challenging and computationally expensive and has not been solved directly through any exact solution method (Chen et al., 2016, Chen et al., 2014, Ekström et al., 2012, Gu et al., 2019). Combining the network level traffic dynamics and simulation-based platforms bring an efficient method to circumvent this issue. Network or macroscopic fundamental diagram (NFD or MFD) represents network traffic flow relationships at the macroscopic scale (Mahmassani et al., 1984, Geroliminis and Daganzo, 2008). The NFD has been widely applied in network-wide traffic control and management (Geroliminis et al., 2012, Haddad, 2017, Haddad et al., 2013, Yang et al., 2017, Keyvan-Ekbatani et al., 2012, Ramezani et al., 2015, Mohajerpoor et al., 2020, Ramezani and Nourinejad, 2018, Han et al., 2020). NFD-based congestion pricing is often conducted by macroscopic modeling of the network with a simulator and the toll level is then adjusted by using the NFD of the pricing zone (Simoni et al., 2015, Zheng et al., 2016, Zheng et al., 2012) which is used as an indicator for monitoring and controlling the congestion level. Zheng et al. (2012) applied a feedback integral control strategy for toll adjustment using an agent-based simulation platform. The integral controller was later extended to a proportional-integral (PI) controller whereby adaptation of road users to pricing was considered (Zheng et al., 2016). A similar simulation framework based on the NFD was also presented by Gu et al. (2018b) including three pricing schemes of distance-based toll, joint distance and time toll (JDTT) and joint distance and delay toll (JDDT). However, they only considered one route choice behavior for loading the network.

In this paper, we consider fixed demand with only two classes of vehicles: HVs that selfishly follow UE strategy and CAVs that seek centrally controlled system optimal routing to enjoy the toll exemption. To capture the distinct microscopic behavior of CAVs, we also implement an adaptive link FD within the SBDTA model. In a heterogeneous network where different road users with different behaviors coexist, a tailored termination criterion is required to ensure the convergence of the mixed equilibrium SBDTA algorithm. Consequently, a hybrid criterion is proposed to find a mixed equilibrium solution in which the travel time experienced by UE-seeking users and the marginal travel time experienced by SO-seeking users departing at the same time between a specific OD pair must be equal and minimal. The second control level in the network is a spatially differentiated distance-based congestion pricing that depends upon the congestion level of the links and the distance traveled within the pricing zone by UE-seeking HVs. We apply a feedback controller to adjust the distance toll rate on each link of the pricing zone with a weight. The weight represents the congestion level of each link implying that high-density links have larger weights whereas low-density links have smaller weights. The so-called spatially differentiated distance-based toll rate can bring more traffic homogeneity within the pricing zone. SO-seeking CAVs are exempted from paying the toll, whereas UE-seeking HVs are subject to the congestion charge.

## 3. Methodology

The methodological framework of this research consists of two components – a mixed equilibrium SBDTA model and a feedback-based spatially differentiated congestion pricing controller.

### 3.1. Mixed equilibrium simulation-based dynamic traffic assignment

We adapt the SBDTA framework originally developed by Mahmassani and Peeta (1993) and Mahmassani (1994) and implement it in AIMSUN, as a network loading engine, to achieve mixed equilibrium accounting for both UE- and SO-seeking CAVs. The three-step framework consists of standard network loading, path set update, and path assignment adjustment, which are applied sequentially and iteratively to all the time-dependent origin-destination (TDOD) demand.

Consider a directed network $G(N, A)$ consisting of a set of nodes $N$ and a set of directed links $A$. Let subscripts 1 and 2 denote UE- and SO-seeking users, respectively. The time horizon of interest is discretized into small time slices $s$, referred to as simulation intervals. Assignment intervals are denoted by time slices $\tau$ where the average travel time and the average marginal travel time are calculated. Let $D$, $S$, and $T$ denote the set of OD pairs, the set of simulation intervals, and the set of assignment intervals, respectively. Hence $q_d^\tau$ represents the demand between OD pair $d \in D$ in assignment interval $\tau \in T$. The total demand of UE- and SO-seeking users are therefore expressed as $q_1^\tau = (\ldots, q_{1d}^\tau, \ldots)$ and $q_2^\tau = (\ldots, q_{2d}^\tau, \ldots)$, respectively, where $q_d^\tau = q_{1d}^\tau + q_{2d}^\tau$ for $\forall d \in D$ and $\forall \tau \in T$.

Fig. 1 illustrates the workflow of the mixed equilibrium SBDTA framework, whose core is a direction-finding mechanism (i.e. obtaining a descent direction for the next iteration) for achieving both UE and SO conditions in the network using results of the current iteration. Time-dependent shortest travel time paths and the least marginal travel time paths are determined in each iteration for UE- and SO-seeking users, respectively and all-or-nothing (AON) assignment is performed. Then the number of SO and UE users in each path are updated by the MSWA. The procedure iterates until a mixed convergence criterion is met.

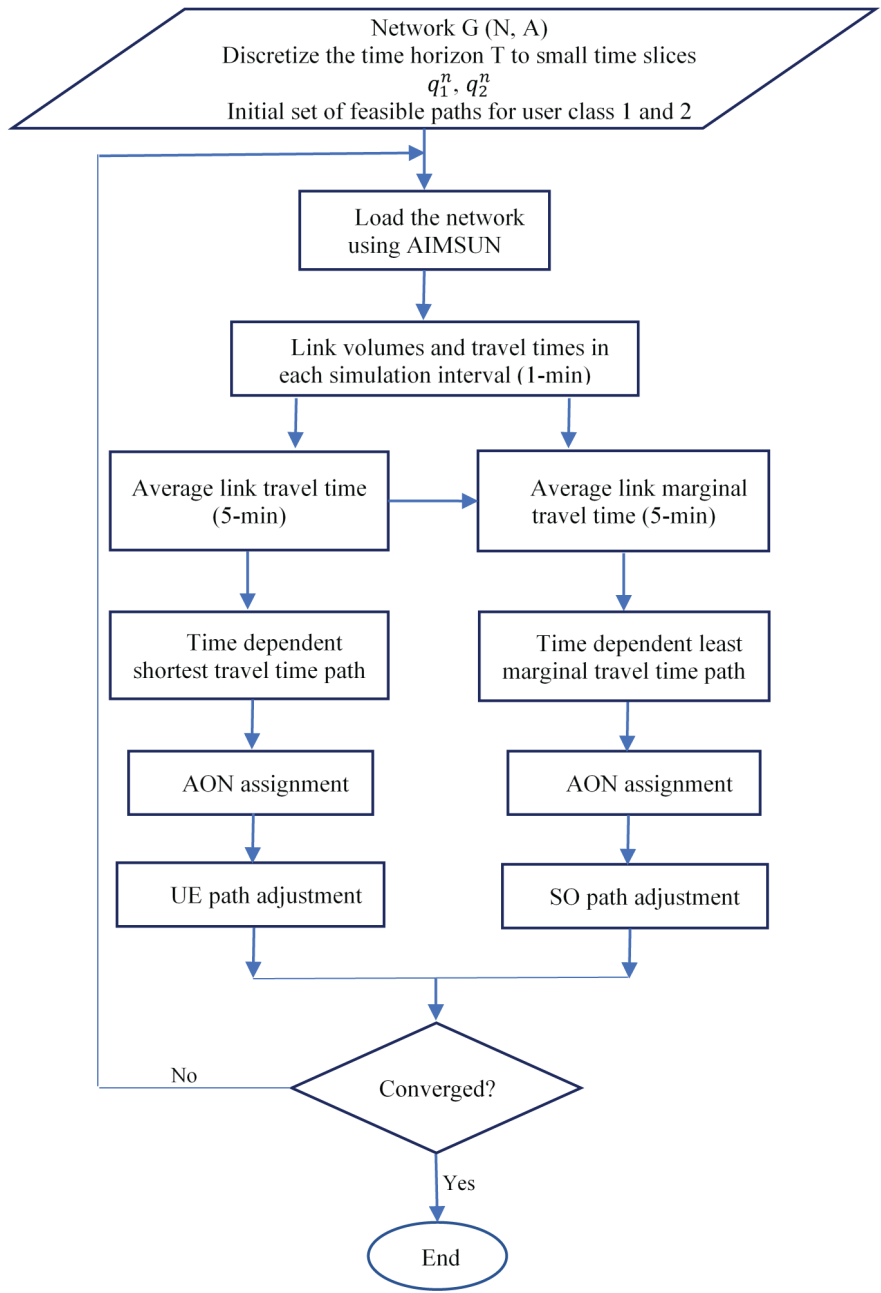

Fig. 1. Adapted mixed equilibrium SBDTA framework based on the original work of Mahmassani and Peeta (1993) and Mahmassani (1994).

### 3.1.1. Calculation of time-dependent marginal travel time

Marginal travel cost is expressed as the derivative of the total system travel time with respect to flow. A global path marginal cost represents the change in the total system travel time caused by an additional vehicle added to the path at a certain time interval. While the global path marginal cost is the most accurate definition, its calculation is computationally expensive due to the need to capture the spatiotemporal interaction of traffic in the network, especially for DTA applications that are time-dependent. In view of this, Ghali and Smith (1995) and Peeta and Mahmassani (1995) proposed a local approximation method to calculate the global path marginal cost. In a nutshell, along a given path, the method first calculates the marginal cost of each link and then sums up all the link marginal costs considering the traversal time of each link. Such a local approximation suggests that the path marginal cost, incurred by an additional vehicle at a certain time interval, only considers the effect of the vehicle on links constituting the path. Although the method may overestimate the path marginal cost due to the time lag between the entry of the additional vehicle to the link and the generated flow perturbation (Qian et al., 2012), it is of practical use especially for large-scale DTA applications.

*3.1.2. Solution algorithm of mixed equilibrium SBDTA*

Many studies in the past have provided practical solution algorithms for the traffic assignment problem (SUE or UE problem) which are based on MSA (Hu et al., 2018, Liu et al., 2009, Powell and Sheffi, 1982). Here we implement a commonly used method in which in each iteration, the OD flow among paths is redistributed using MSA and the demand percentages that are to assign to each route are updated. Let predetermined sequence of step size in MSA $\theta_n = \frac{1}{n}$, where $n$ is the iteration number ($n = 1, 2, ...$) and $(y^n - x^n)$ is auxiliary and current flow difference as the descent direction in search process. The conventional MSA is calculated as

$$x^{n+1} = x^n + \theta_n(y^n - x^n) \tag{1}$$

that can also be rewritten as

$$x^{n+1} = \frac{1}{n}(y^1 + y^2 + \cdots + y^n) \tag{2}$$

At the beginning of the MSA process, $\theta_n$ could be too large, and therefore, the objective values do not decrease until several iterations. On the other side, after many iterations, the step size could become too small leading to slow convergence speed. As SBDTA solution algorithm lacks link cost and link exit functions, it is required to have alternative step size choice strategies to improve the performance of MSA and maintain its simplicity. In this study, MSWA is examined to give higher weights to the later auxiliary flow patterns. Considering that the summation of all weights should be equal to 1, a generalized formulation of MSWA can be expressed as (Liu et al., 2009)

$$x^{n+1} = \frac{1^\gamma \times y^1 + 2^\gamma \times y^2 + 3^\gamma \times y^3 + \cdots + n^\gamma \times y^n}{1^\gamma + 2^\gamma + 3^\gamma + \cdots + n^\gamma} \tag{3}$$

that can also be rewritten as

$$x^{n+1} = x^n + \frac{n^\gamma}{1^\gamma + 2^\gamma + 3^\gamma + \cdots + n^\gamma}(y^n - x^n) \tag{4}$$

where $\gamma$ is weight parameter here which is a real number and determines how much weight is considered for the later iterations. It is obvious that the conventional MSA is a special case of MSWA when *γ=0*. The step size for different weight parameters are as follows:

$$\theta_n = \frac{1}{n} \quad (\gamma=0) \tag{5}$$

$$\theta_n = \frac{2}{n+2} \quad (\gamma=1) \tag{6}$$

$$\theta_n = \frac{6n}{(n+1)(2n+1)} \quad (\gamma=2) \tag{7}$$

The detailed algorithmic steps of the implemented mixed equilibrium SBDTA are summarized in Table 1. Existing approaches in the literature to solving the equilibrium problem in large-scale networks that adopt a simulator for network loading are heuristic and thus, no mathematical proof of convergence can be provided to determine an exact solution. For example, approximate solution of dynamic user equilibrium can be reached when the experienced travel time between a specific OD pair and departure time are equal and minimal reflected in a proposed relative gap function by (Janson, 1991). In a mixed traffic network where different road users with different behaviors coexist at the same time, a new convergence criterion is required. In this study a hybrid criterion is proposed to find a mixed equilibrium solution in which the travel time experienced by UE-seeking users and the marginal travel time experienced by SO-seeking users departing at the same assignment time interval between the same OD pair are equal and minimal. The proposed SBDTA framework is considered being converged if the average value of UE and SO equilibrium relative gap (equation (8) and (9), respectively) become stable and less than $\varepsilon_1$.

$$R_1 gap(i) = \frac{\sum_{\tau \in T} \sum_{d \in D} \sum_{p \in P_d} f^i_{1,d,p}(\tau)[tt^i_{d,p}(\tau) - \pi^i_d(\tau)]}{\sum_{\tau \in T} \sum_{d \in D} q_{1,d}(\tau) \pi^i_d(\tau)} \quad (8)$$

$$R_2 gap(i) = \frac{\sum_{\tau \in T} \sum_{d \in D} \sum_{p \in P_d} f^i_{2,d,p}(\tau)[mt^i_{d,p}(\tau) - \delta^i_d(\tau)]}{\sum_{\tau \in T} \sum_{d \in D} q_{2,d}(\tau) \delta^i_d(\tau)} \quad (9)$$

where $f^i_{1,d,p}(\tau)$ and $f^i_{2,d,p}(\tau)$ is the flow of UE and SO-seeking users on path $p$ between OD pair $d$ at time interval $\tau$ at iteration $i$, $q_{1,d}(\tau)$ and $q_{2,d}(\tau)$ is the demand of UE- and SO-seeking users between OD pair $d$, $tt^i_{d,p}(\tau)$ and $mt^i_{d,p}(\tau)$ is the experienced travel time and marginal travel time of path $p$ between OD pair $d$, $\pi^i_d(\tau)$ and $\delta^i_d(\tau)$ is the least travel time and marginal travel time path between OD pair $d$ at time $\tau$ at iteration $i$.

Table 1: Mixed equilibrium SBDTA algorithm

**Input:** network $G(N,A)$, demand $q_1^\tau$ and $q_2^\tau$, gap tolerances $\varepsilon_1$, and the maximum numbers of iterations N.
**Output:** optimal paths choices of UE- and SO-seeking users $p_1^*$ and $p_2^*$ under mixed equilibrium
**Initialization**
Set the iteration index $i = 1$, obtain distance-based shortest path between for every OD pairs and assign the demand $q_1^\tau$ and $q_2^\tau$ to it for each $\tau \in T$.
**Main loop**
    **While** $i < N$ and $Rgap(i) > \varepsilon_1$
      i. Load the network using AIMSUN
      ii. Find shortest paths
          -Find the time-dependent shortest path for each OD pair and each assignment interval; perform AON for UE-seeking HVs and obtain the auxiliary matrix $y_1^i$.
          -Find the time-dependent least marginal travel time path for each OD pair and each assignment interval; perform AON for SO-seeking CAVs and obtain the auxiliary matrix $y_2^i$.
      iii. Update paths choices via MSWA
$$p_1^{i+1} = p_1^i + \alpha_1^i * (y_1^i - p_1^i) \quad (10)$$
$$p_2^{i+1} = p_2^i + \alpha_2^i * (y_2^i - p_2^i) \quad (11)$$

      iv. Calculate path flows.
$$f_1^{i+1} = p_1^{i+1} * (q_1^\tau)^T \quad (12)$$
$$f_2^{i+1} = p_2^{i+1} * (q_2^\tau)^T \quad (13)$$

      v. Check convergence.
$$Rgap(i) = \frac{R_1 gap(i) + R_2 gap(i)}{2} \quad (14)$$
      vi. $i = i + 1$
    **End while**

*3.1.3. Adaptive link fundamental diagram*

The CAVs are expected to have the capability of driving in platoons at smaller headways compared to HVs. Therefore, a finite number of CAVs utilizes a smaller space on the road compared to HVs. Also, CAVs' capability in responding to and absorbing traffic flow fluctuations could result in a larger critical density, where the capacity drop occurs, in a link (Levin and Boyles, 2016, Simoni et al., 2019). To capture this distinct microscopic behavior in the mixed autonomy traffic, we implement "adaptive link fundamental diagram" within the developed SBDTA framework that allows varying reaction times and capacities depending on the fraction of CAVs on a link during the simulation.

To capture the impact of mixed autonomy traffic with different reaction times on link capacity and the shockwave speed, we compute an average link reaction time in every assignment interval, based on the percentage of CAVs and HVs entering each link in the previous time interval. A shorter reaction time is considered for CAVs (1 second) compared to the HVs (1.5 second) (Fakhrmoosavi et al., 2020, Levin and Boyles, 2016) to produce the desired adaptive link fundamental diagrams. As an example, Fig. 2Fig. 2 illustrates the triangular flow-density relationship for a link with the speed limit of 60 km/hr and vehicles with effective length of 7 m (sum of the vehicle length and the minimum distance between vehicles) (Tss, 2014). As the average link reaction time reduces, the critical density and the shockwave speed increase and shifts the peak of the fundamental diagram towards the right. A further discussion on the simulated link flow-density relationships is provided in Appendix A.

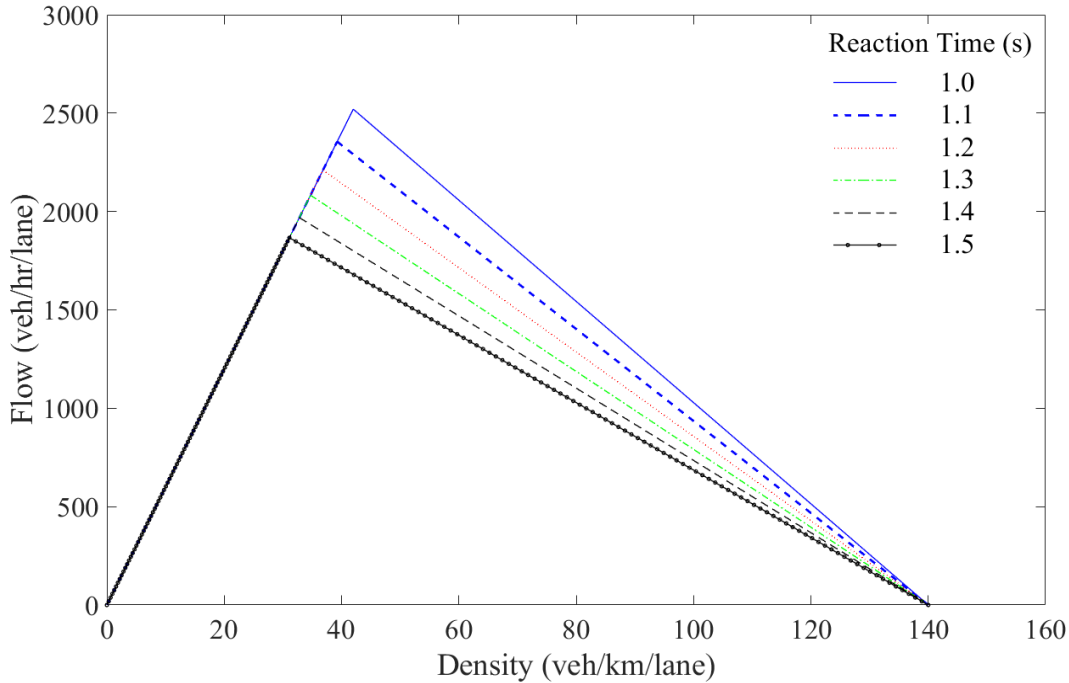

Fig. 2. Link flow-density relationship as a function of average link reaction time.

### 3.2. Feedback-based congestion pricing controller

Given the potential increase in travel demand because of the presence of CAVs and the accordingly comfort they may provide (Litman, 2017), congestion pricing as an effective travel demand management strategy is likely to become more justifiable from a political and public acceptance perspective. The proposed pricing scheme in this paper does not penalize HVs, rather they provide incentives to CAVs who are willing to give away their selfish routing (UE) to a centrally controlled routing system (SO). In a congested network, therefore, we expect that SO-seeking CAVs whose routing is being controlled centrally may have to take longer routes. In return, they will receive exemption from paying the congestion toll.

The proposed joint routing and pricing scheme seeks two aims. The first is to encourage and incentivize CAVs to follow SO strategy given their routing controllability from a technological point of view. The second aim is to discourage UE-seeking HVs from congesting the pricing zone by imposing a toll. HVs, therefore, have two options: (1) make a detour and avoid entering the pricing zone and thus, avoid paying the toll and (2) continue with their original route through the pricing area and pay the congestion toll.

The distance traveled within the pricing zone is the main determinant in the developed pricing scheme. Road users are charged relative to their distance traveled within the pricing zone. This scheme is believed to be more equitable compared to the classical cordon based pricing (e.g. pay-per-entry) (Daganzo and Lehe, 2015). However, distance-based toll tends to force users to take the shortest paths within the pricing area to pay less. As a result, travelers may drive on the same shortest paths leading to more heterogeneous congestion within the pricing area that exacerbates the hysteresis loop in the NFD (Gu et al., 2018b, Gu and Saberi, 2021). To enjoy distance-based pricing benefits and create a more homogenous distribution of traffic congestion, we differentiate the distance-based toll rates by the congestion level on each link. To do so, we apply a feedback controller to adjust the distance toll rate on each link with a weight. The weights represent the congestion level of each link in which high-density links have larger weights whereas low-density links have smaller weights. The so-called spatially differentiated distance-based toll rate is expected to make the congestion distribution more homogeneous within the pricing area and help to reduce the size of the NFD hysteresis.

Table 2: Summary of notations

| Notation | Definition |
|---|---|
| $n_a$ | Number of lanes of link $a$ |
| $l_a$ | The length of link $a$ |
| $\omega$ | Weight parameter associated with congestion level of each link |
| $k_a(\tau)$ | The average density of link a during interval $\tau$ |
| $q_a(\tau)$ | The average flow of link a during interval $\tau$ |
| $\delta_{a,p}^d$ | $\delta_{a,p}^d = 1$ if path $p \epsilon P^d$ contains link $a$, otherwise $\delta_{a,p}^d = 0$ |
| $l_p^d(\tau)$ | Total distance traveled within subnetwork $\bar{G}$ for path $p \epsilon P^d$ and interval $\tau$ |
| $\alpha(\tau)$ | Distance-based toll rate during interval $\tau$ ($/km) |
| $\bar{\alpha}_p^d(\tau)$ | Distance-based toll component for path $p \epsilon P^d$ and interval $\tau$ |

Consider a subnetwork $\bar{G}(\bar{N}, \bar{A})$ of the initial network which is set as the pricing zone. see Table 2 for the notations. We assume that the spatially differentiated distance-based toll function $\emptyset^\tau(d, \omega)$ during interval $\tau$ is related to the distance traveled by the UE-seeking users in each link and a weight parameter associated with its congestion level. Note that $\alpha(\tau)$ is a positive distance-based toll rate that is determined using a feedback control strategy for every assignment interval which will be discussed later.

$$\emptyset^\tau(d, \omega) = \alpha(\tau).(1 + \omega).d \tag{15}$$

The congestion level is determined as a relative change of link actual and free flow travel time as expressed in Equation 16. This parameter can take a number from 0 in free flow state to a considerably high value in heavily congested state. The maximum value of the ω should be adjusted in real-world applications which could be predefined by the transport authority. In this study, ω changes between 0 to 1 meaning that the distance toll rate in congested links can be adjusted up to twice as high as uncongested links.

$$\omega = \frac{t_a(\tau) - \bar{t}_a}{\bar{t}_a} \tag{16}$$

We obtain the spatially differentiated distance-based toll component for path $p \epsilon P^d$ during interval $\tau$ as follows

$$\bar{\alpha}_p^d(\tau) = \alpha(\tau).\sum_{a \in \bar{A}}(1 + \omega).l_a.\delta_{a,p}^d \tag{17}$$

When $\omega = 0$, the spatially differentiated distance toll component reduces to the original distance-based toll (Gu et al., 2018b). Given the spatially differentiated distance-based toll component, the generalized cost function of path $p \epsilon P^d$ during interval $\tau$ for the UE-seeking users $U_{p,1}^d(\tau)$ and the SO-seeking users $U_{p,2}^d(\tau)$ are respectively expressed as follows

$$U_{p,1}^d(\tau) = \sum_{a \in A} tt_a(\tau).\delta_{a,p}^d + \frac{\bar{\alpha}_p^d(\tau)}{VOT} \tag{18}$$

$$U_{p,2}^d(\tau) = \sum_{a \in A} mt_a(\tau) \cdot \delta_{a,p}^d \tag{19}$$

The generalized costs of paths between each O-D pair are used in the mixed equilibrium algorithm to achieve the mixed equilibrium flow under the pricing control.

A feedback control strategy is applied to adjust the distance-based toll rates for every toll interval. This approach has wide applications in traffic management and control, including ramp metering (Papageorgiou et al., 1991), perimeter or gating control (Keyvan-Ekbatani et al., 2012, Aboudolas and Geroliminis, 2013, Ramezani et al., 2015), and congestion pricing (Gu et al., 2018b, Zheng et al., 2012, Zheng et al., 2016). It works by iteratively adjusting the control input based on the feedback output to achieve a predetermined setpoint. When integrating feedback control with the NFD for congestion pricing applications, the objective is typically to pricing-control the network such that the NFD does not enter the congestion regime and operates instead around the critical network density.

The overall process of the feedback-based congestion pricing integrated with mixed equilibrium SBDTA is illustrated in Fig. 3. Following Zheng et al. (2016) and Gu et al. (2018b), a discrete proportional-integral (PI) controller is employed as expressed below

$$\alpha_\tau(i) = \begin{cases} \alpha_\tau(i-1) + P_P\big(\overline{K}_\tau(i) - \overline{K}_\tau(i-1)\big) + P_I(\overline{K}_\tau(i) - K_{cr}), & i > 1 \\ P_I(\overline{K}_\tau(i) - K_{cr}), & i = 1 \end{cases} \tag{20}$$

where $\alpha_\tau(i)$ is the toll rate at tolling interval $\tau$ (which is consistent with assignment interval) at iteration $i$, $\overline{K}_\tau(i)$ is the average network density within the $\tau$-th tolling interval during iteration $i$, $P_P$ and $P_I$ are proportional and integral gain parameters to be tuned via trial and error, and $K_{cr}$ is the critical network density identified from the NFD. A sensitivity analysis on $P_p$ and $P_I$ has already been performed on the same network in Gu et al. (2018b). Therefore, in this study, the same values for the gain parameters are used as in Gu et al. (2018b). To calculate average network density $\overline{K}_\tau$ and average network flow $\overline{Q}_\tau$ in every tolling interval $\tau$, the outputs of our mixed equilibrium SBDTA are extracted and substituted in the following equations (Saberi et al., 2014)

$$\overline{K}_\tau = \frac{\sum_{a \in A} k_a(\tau) \cdot l_a \cdot n_a}{\sum_{a \in A} l_a \cdot n_a} \tag{21}$$

$$\overline{Q}_\tau = \frac{\sum_{a \in A} q_a(\tau) \cdot l_a \cdot n_a}{\sum_{a \in A} l_a \cdot n_a} \tag{22}$$

The overall workflow of Fig. 3 is summarized in the following steps:

- $K_{cr}$ is identified from the NFD when no toll is imposed on the UE-seeking users in the pricing zone and is considered as the setpoint in the PI controller.
- The discrete PI controller (Equation (20)) tries to drive the average network density towards the critical density by adjusting the toll rates.
- Considering non-uniform ratios of SO and UE-seeking users and adjusted toll rate vector in the previous step, mixed equilibrium SBDTA algorithm (Table 1) is executed to determine the average density of the pricing zone.

- The elements of the PI controller are updated, and the cycle is run until the average density of pricing zone gets sufficiently close to the critical density subject to the maximum toll that the system can impose.

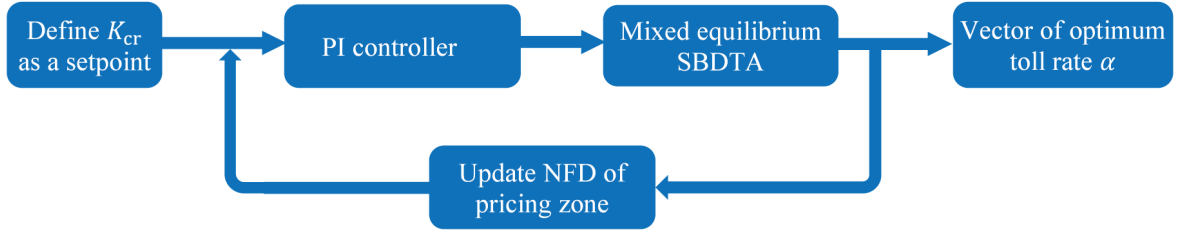

Fig. 3. Feedback-based congestion pricing approach.

### 3.3. Mathematical formulation of the joint routing and pricing control in the mixed traffic flow of CAVs and HVs

The proposed joint routing and pricing control scheme in the mixed traffic flow falls into the broad category of transportation network design problems (Sheffi and Powell, 1983, Florian and Chen, 1995, Yang, 1995). These problems are commonly formulated as a bi-level problem with decision variables being optimized in the upper level and the lower level is a traffic assignment problem considering the optimized decision variables. In our proposed modeling framework, the decision variables are grouped into a vector consisting of the distance toll rates for every 15-minute period. We assume the VOT is equal to 15 $/h (Legaspi and Douglas, 2015).

Consider the following equivalent mathematical formulation of the NFD-based toll rate problem.

$$min_{[\alpha_\tau]} \sum_{\tau \in T} |\bar{k}_\tau - k_{cr}| \qquad (23)$$

subject to

$$0 < \alpha_\tau \leqslant \bar{\alpha} \qquad (24)$$

$$\bar{k}_\tau = \text{mixed equilibrium SBDTA}(\alpha_\tau) \qquad (25)$$

The upper-level objective function, Equation (23), seeks to calculate the toll vector within the tolling period in order to maintain the average density of the pricing zone sufficiently close to the critical density identified from the initial NFD of the pricing zone. Constraint (24) bounds the feasible regions for adjusting the toll rate where $\bar{\alpha}$ is the upper bounds specified by transport authorities. Constraint (25) represents the lower-level mixed equilibrium SBDTA considering the control inputs $\alpha_\tau$ to determine the average density of the pricing zone. The simulation output $\bar{k}_\tau$ is fed back to the optimization problem and the process keeps continuing iteratively until a certain termination criterion is met.

We investigate different exogenously given ratios of SO-seeking CAVs, ranging from 0% to 100%, that are non-uniformly distributed across the network. To allow heterogeneous distribution of SO-seeking CAVs across the network, we introduce random variations (noise) to the uniform ratio of SO-seeking users in order to obtain a non-uniform distribution (Saffari et al., 2020). We assume that the random noise follows a uniform distribution, $\varepsilon \sim U(-\beta \cdot q_d^\tau, \beta \cdot q_d^\tau)$, where $\beta \sim U(0, 0.2)$ and $q_d^\tau$ denotes the total demand between OD pair $d \in D$ in assignment interval $\tau \in T$. For each scenario, we generate separate random noise levels for each OD pair in each assignment interval which are added to the predefined uniform ratios of SO-seeking users, $q_{2,d}^\tau + \varepsilon$.

## 4. Numerical results and discussion

### 4.1. Application of the mixed equilibrium SBDTA to a small network without pricing

In this section, we aim to explore the convergence pattern of the proposed solution algorithm considering MSA and MSWA (with different weights) used in the path assignment adjustment step. The section also provides the numerical results of the mixed equilibrium SBDTA framework with different proportions of SO-seeking CAVs in a small network to examine and verify its performance before applying to a large-scale network. To do this, Nguyen network is used consisting of four O–D pairs, 19 links and 25 routes. The simulation includes 12 assignment intervals with 5-min duration and the last nine of which have zero demand to fully empty the network to facilitate a reasonable comparison of the total system travel time. Note that in this section, the proportion of SO-seeking CAVs is uniformly distributed across all the OD pairs in the assignment intervals.

Fig. 5 demonstrates that MSA performs better than MSWA when the number of iterations performed is relatively low, however, it is unable to achieve significantly reduced relative gap when the number of iterations further increases. Convergence speed of MSWA with a higher weight parameter (e.g. $\gamma=2$) becomes faster when the number of iterations increases. This means that if a small relative gap is not required because of the simulation time and computational limitation particularly for a large-scale network (without consideration of parallel computing approaches (Chen et al., 2020)), the MSA performance is still desirable. As shown and compared in Fig. 6, the total system travel time in the mixed equilibrium model decreases, as expected, with an increasing proportion of SO-seeking CAVs in the network. Compared to the full UE scenario, deploying 100% SO-seeking CAVs can effectively reduce the total system travel time by 7% (from 183 hr to 170 hr). It also shows that the network performance can be improved significantly by deploying 20% SO-seeking CAVs. The improvement is relatively small, however, under 40% and 60% SO ratio scenarios, but it becomes significant again once the SO ratio in the network exceeds 60%.

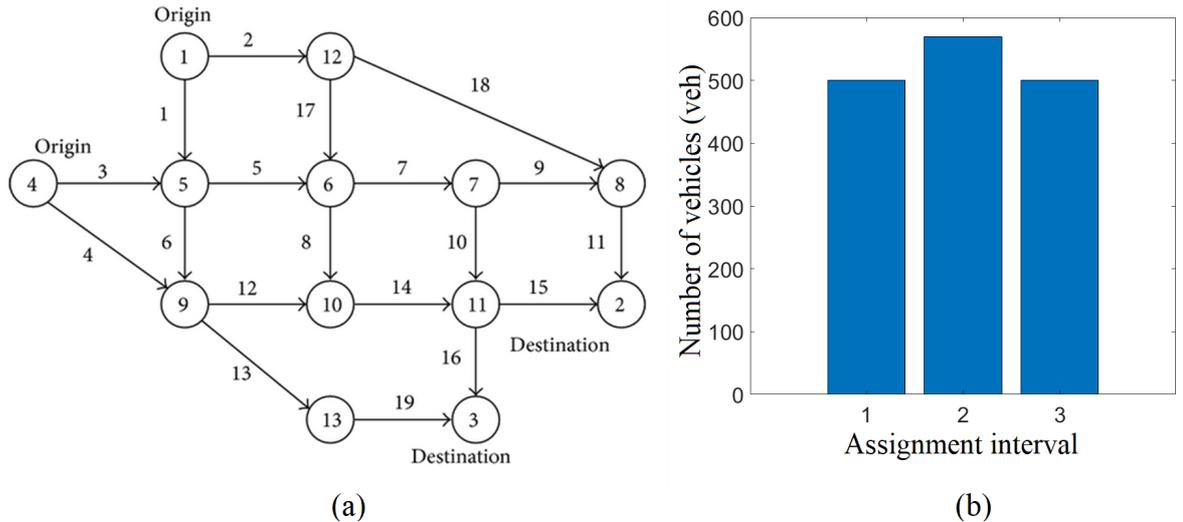

(a)                              (b)

Fig. 4. Topology and demand profile of the Nguyen network.

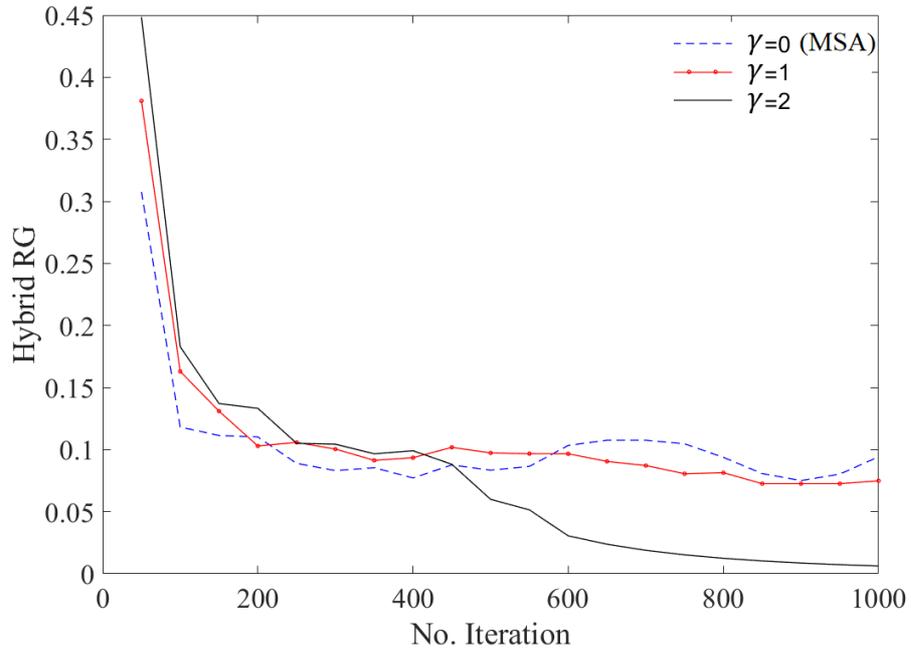

Fig. 5. Comparison of convergence trend of MSA and MSWA applied to the Nguyen network.

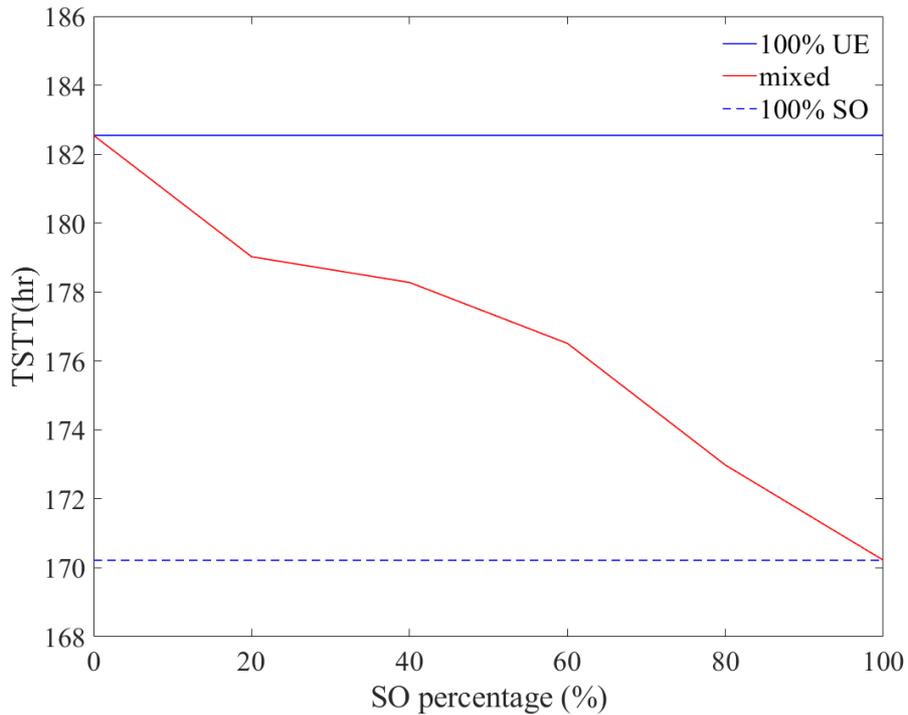

Fig. 6. Total system travel time (TSTT) under different ratios of SO-seeking CAVs in the Nguyen network.

### 4.2. Application of the proposed routing and pricing schemes to a large network

A comprehensive comparison is performed between two strategies consisting of routing control only and joint routing and pricing control. The proposed methodology is applied to a large-scale network model of Melbourne, Australia (Shafiei et al., 2018). The model is deployed in AIMSUN, as the mesoscopic simulator engine, with time-dependent travel demand from 6 to 10 AM.

*4.2.1. Numerical results of the routing only control strategy*

This section aims to explore the efficiency and computational performance of the mixed equilibrium SBDTA framework in a large-scale traffic network. A subnetwork is first extracted from the greater Melbourne area model including a pricing zone in the center (See Fig. 7(a)). The area of the subnetwork is about 140 square kilometer in which UE-users will be tolled if they enter the green colored rectangular zone in the center. The network topology and the distribution of the time-dependent OD demand are shown in Fig. 7 and summarized in Table 3, respectively. The number of allocated OD routes to SO-seeking CAVs is dictated by a central agent that may be varying in different time intervals. The number of assigned paths to SO-seeking users by the central agent for a given OD pair and assignment interval could be flexible, whereas the path set size for UE-seeking users are predefined to correspond the idea of well-known paths for the drivers. It is assumed that the path set size of SO- and UE-seeking users are at most 5 and 3 paths, respectively.

Table 3: Network topology of the extracted sub-network

| | |
|---|---|
| Area of the network | 9.5 x 14.6 (sq.km) |
| Total number of links | 4,375 |
| Total number of nodes | 1,977 |
| Total number of centroids | 492 |
| Area of the pricing zone | 1 x 2 (sq. km) |
| Number of links in the pricing zone | 282 |
| Number of nodes in the pricing zone | 91 |
| Number of centroids in the pricing zone | 30 |

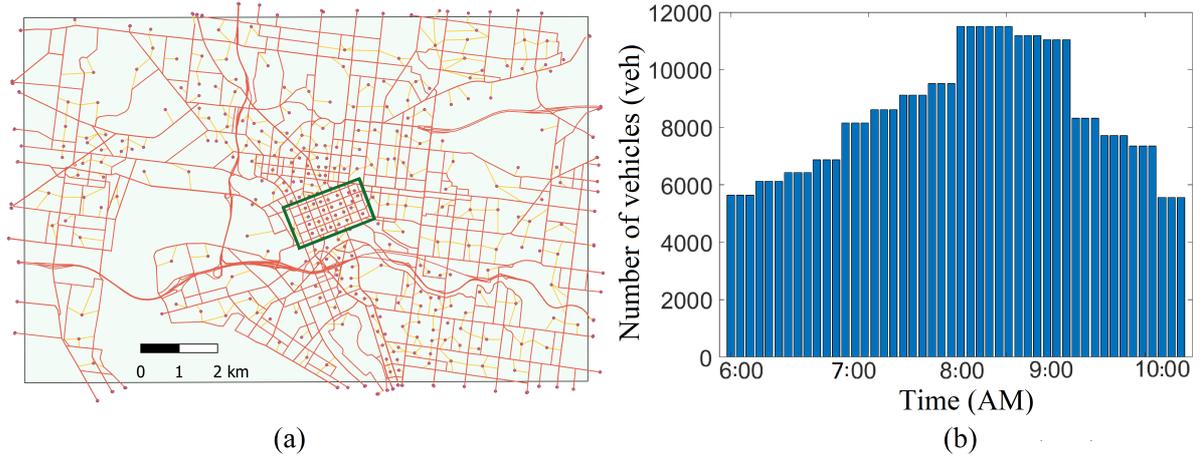

(a)          (b)

Fig. 7. The extracted subnetwork from the Greater Melbourne area model; (b) time-dependent demand profile.

As shown in Fig. 8, the mixed equilibrium SBDTA model can be considered converged within 40 iterations. Fig. 9 illustrates the NFD of the CBD (green rectangle in Fig. 7 which is later considered as the priced zone once we implement the pricing control) and the whole network with its loading and unloading phases under full routing control (100% SO-seeking CAVs) and non-controlled (100% UE-seeking users) scenarios.

As shown in Fig. 9 (a), the CBD enters the saturated and over saturated regime in the non-controlled scenario in which the critical density is determined as 15 veh/km and the maximum network flow is 520 veh/hr. While under full routing control scenario, the network mostly performs in free flow and the maximum network flow is 600 veh/hr, slightly larger than 100% UE scenario. Results suggest that employing CAVs with routing capabilities and unique microscopic behaviors will likely yield improvements in the experienced maximum flow in the CBD and the whole of the network.

In the recovery phase, the distribution of traffic congestion tends to be more heterogeneous as the congested spots dissipate more slowly than the low congested areas. This spatially heterogeneous congestion can reduce the network flow throughput during unloading leading to a clockwise hysteresis loop in the NFD (Gayah and Daganzo, 2011). The hysteresis loop in the 100% UE scenario is larger compared to 100% SO scenario. By increasing the proportion of SO-seeking users in the network, the distribution of traffic in the network becomes more homogenous and thus, the size of the hysteresis loop reduces. In a network with high proportion of SO-seeking users, traffic congestion also tends to dissipate more rapidly in the unloading phase.

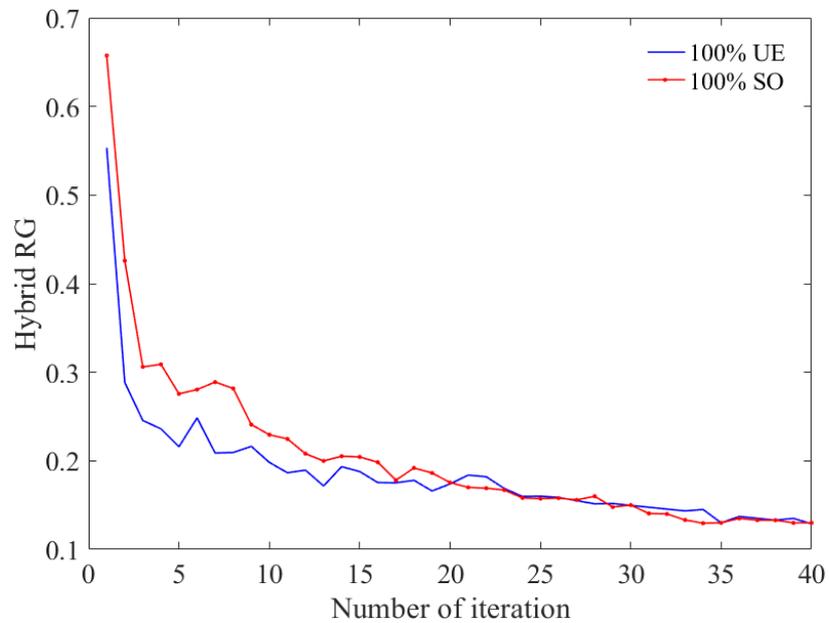

Fig. 8. Convergence pattern of the implemented solution algorithm for the Melbourne network using the proposed hybrid relative gap.

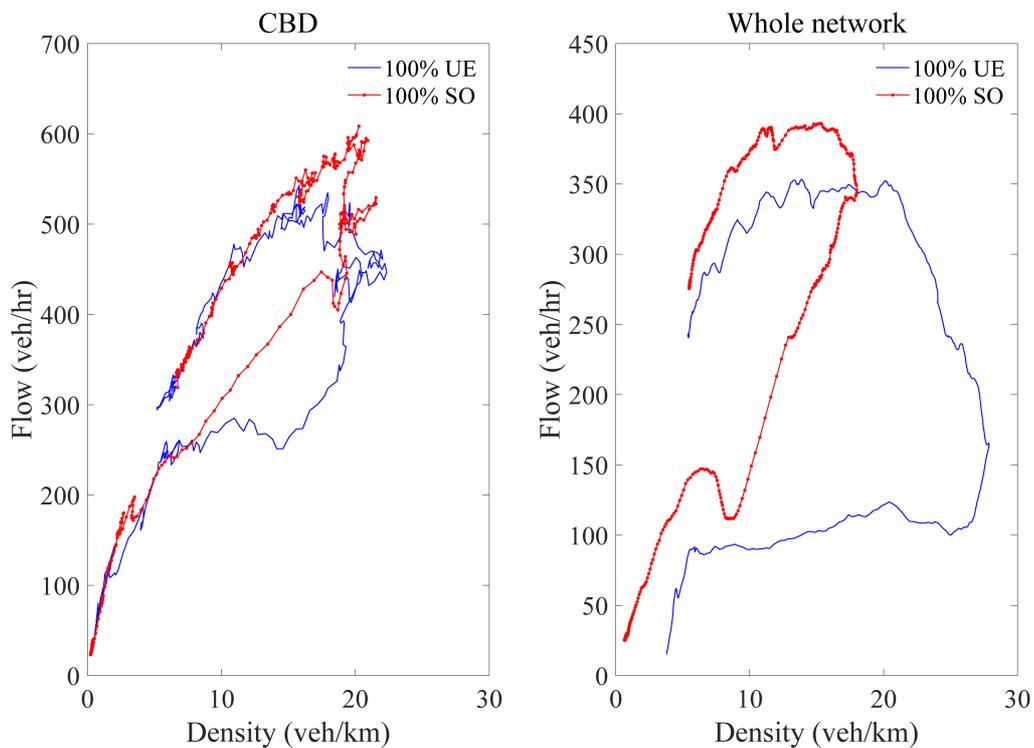

Fig. 9. NFD of the CBD and the whole network for 100% UE and 100% SO scenarios.

However, Fig. 10 reveals that the overall traffic density in the CBD (with no pricing) has not been influenced considerably even when the routing of all vehicles in the network is controlled. Under full routing control scenario in which all the vehicles are SO-seeking, although the capacity of the CBD is improved, its density reaches close to 25 veh/km in the morning peak hour. Although increasing the proportion of SO-seeking CAVs will reduce the system travel time, it does not necessarily improve traffic density in the CBD area and thus, further traffic management policies are required to relieve congestion. Hence, the potential impact of congestion pricing in a mixed autonomy network is worthy of investigation and is implemented for the time periods when the average density in the CBD exceeds 15 veh/km (i.e. simulation period of 120-240 min).

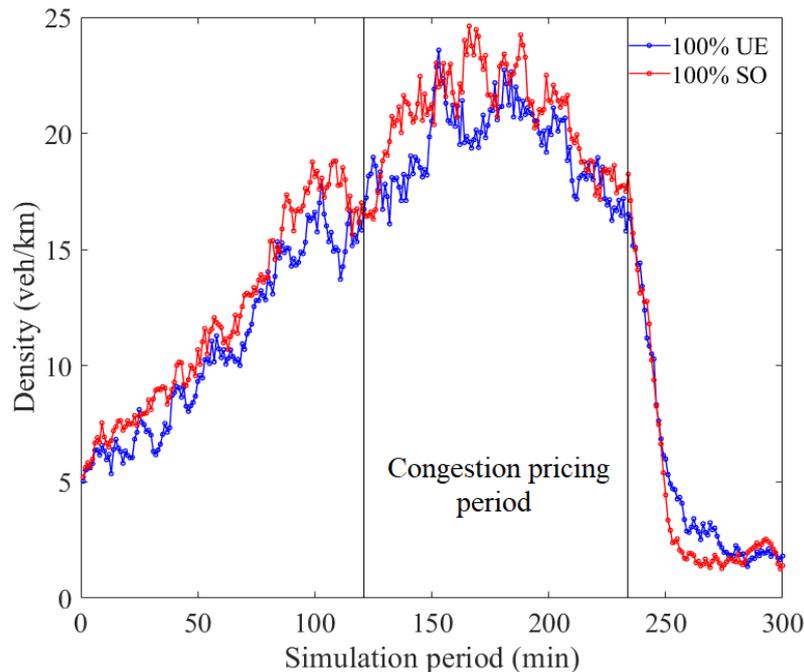

Fig. 10. Density time-series of Melbourne CBD subnetwork for 100% UE and 100% SO scenarios.

### 4.2.2. Numerical results of joint routing and pricing control

In this section, the network performance under joint routing and pricing control is analyzed for different proportions of SO-seeking CAVs in the network. Time dependent and spatially differentiated distance toll is imposed on the UE-seeking users while the CAVs that allow themselves to be controlled by the central agent, are exempted from paying the toll for travelling within the pricing zone. Obviously, having more SO-seeking users in the network reduces total system travel time but the CBD area will not necessarily have less congestion. On the other hand, the toll rate could be set unfairly high for the remaining UE users to prevent them from entering the pricing zone. Therefore, an appropriate combination of SO- and UE-seeking users should exist that not only reduces the TSTT significantly across the entire network, but also keeps the CBD uncongested by imposing a fair toll rate on HVs. Here, the NFD-based toll rate is estimated for 5 scenarios with different ratios of SO-seeking CAVs in the network. Fig. 11(a to e) show the simulation results of the pricing zone when the optimal spatially differentiated distance toll is imposed. Results suggest that by implementing a dynamic distance toll, the traffic congestion in the pricing zone is relieved, as expected.

In the 100% UE scenario, the distance toll rate is minimum compared to the other scenarios. This rate in the third toll interval [8:30 - 8:45] reaches its highest value (0.27 $/km) to achieve the control objective and to reduce the average pricing zone density from 20 to 15veh/km. By increasing the penetration rate of SO-seeking users in the network, the toll rate increases to

keep the average density of the pricing zone below the critical density. In the 60% SO scenario, an average density of 16.8 veh/km is achieved with an average toll of 0.97 $/km for the pricing zone. When there are only 20% UE-seeking users in the network who are subject to tolls, an average density of 18.68 veh/km is achieved under an average toll of 1.71 $/km. As previously mentioned, the proposed spatially differentiated distance-based pricing differentiates toll rates on the links to encourage HVs to use less congested links. This is expected to create a more homogenous distribution of congestion in the pricing area. Fig. 11 (f to j) illustrates the distribution of the link tolls across the pricing zone in different scenarios. Generally, the congestion level of two-thirds of the links is considerably high, while the rest of the links are less congested. Results suggest that the more SO users in the network, the higher toll rates are imposed on the remainder of UE users especially if they select the congested links within the pricing zone. The toll difference between the congested and uncongested links tends to provide a more affordable option for UE-seeking users to switch and select the uncongested links to improve the spatial distribution of congestion. The spread of pricing rates among links directly depends on the congestion level of the links, ω. As previously defined, ω can take a number from 0 (in free flow state) to 1 (in heavily congested state) meaning that the distance toll rate in congested links can be adjusted up to twice as high as uncongested links.

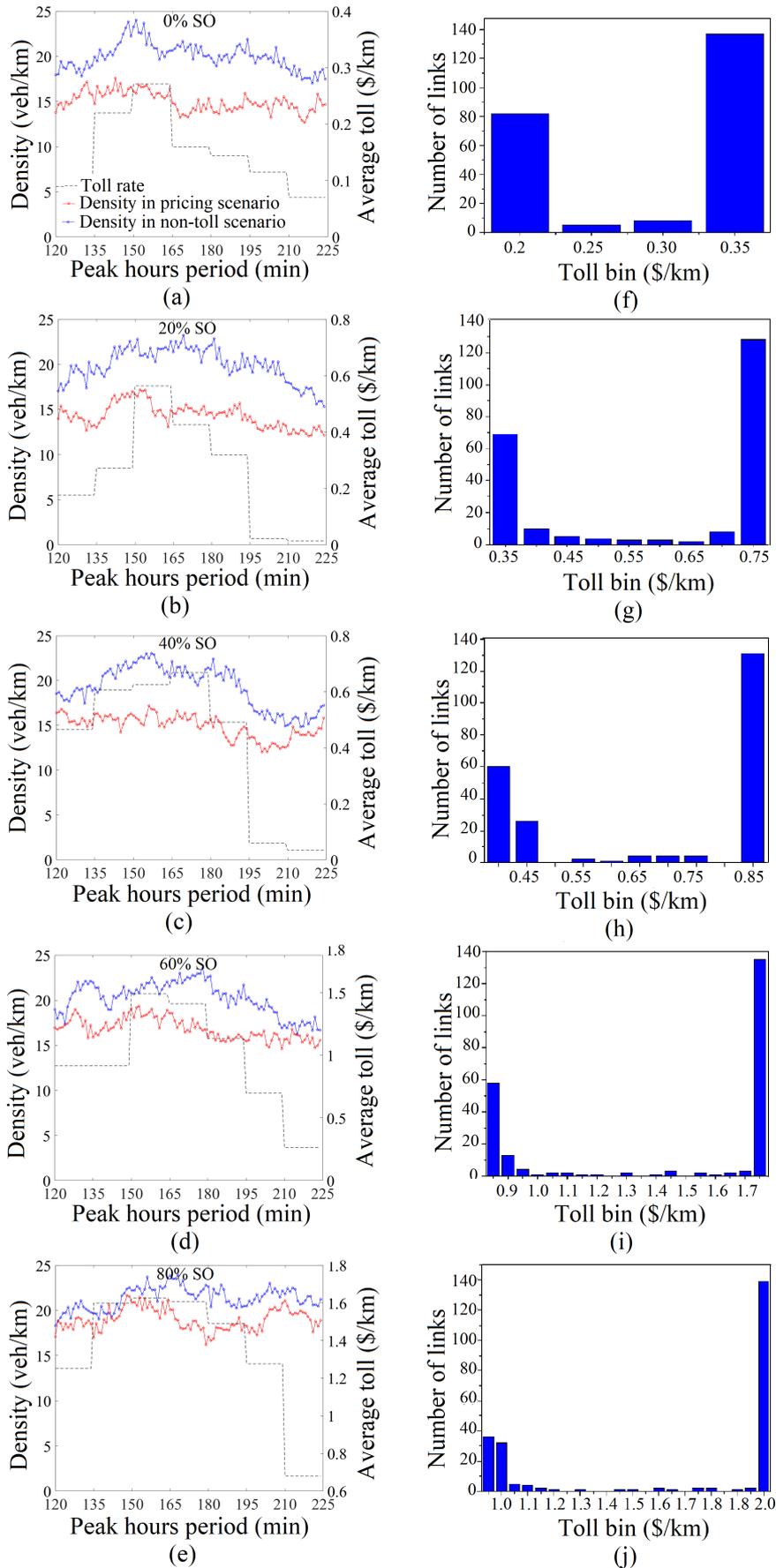

Fig. 11. Density time-series of the pricing zone before and after imposing the toll pricing for different SO ratios (a-e) along with the presentation of link toll distribution in 8:30 AM to 8:45 AM interval (f-j).

The pricing area experiences oversaturated condition in low SO ratios. Increasing the ratio of SO users (more than 40% in our case) can prevent the network from entering the congested branch of the NFD. This improvement is also reflected in the experienced maximum flow in the pricing area that has increased from 500 veh/hr (in 100% UE non-toll scenario) to 600 veh/hr (in 100% SO non-toll scenario).

Generally, implementing congestion pricing improves the performance of the pricing zone in all scenarios. The more UE-seeking users in the network, the more potential vehicles are available to detour and avoid entering the pricing area (e.g., 0 and 20% SO toll-scenarios). So, traffic condition will remain uncongested where the average flow and density is almost 500 veh/hr and 15 veh/km, respectively. Although a moderate or high penetration rate of SO-seeking users increases the maximum flow in the pricing area (550 veh/km), average density increases and reaches 17-18 veh/km (still in the uncongested branch of the NFD, though).

An interesting observation here is the shape of the hysteresis loop in different scenarios. When a single route choice behavior is dominant in the network, the size of the hysteresis loop becomes quite large. The size of the hysteresis loop is considerably large when there is no vehicle following SO routing. When SO ratio reaches 100% the hysteresis loop still appears but with a smaller size compared to the 100% UE scenario. While the selfish UE-seeking users seek to minimize their own travel time and therefore, take the path with the shortest travel time which leads to a more heterogeneous congestion distribution, the number of allocated routes to SO-seeking CAVs by the central agency is more than the number of routes that UE-seeking users can adopt in each time interval. This results in a more homogeneous distribution of SO-seeking users in the network. Mixed scenarios including 40 and 60% SO users have the largest impact on the hysteresis loop. The main reason for such behavior is that in the case of high ratio of UE or SO users, most of the feasible paths for UE users (paths with shortest travel time) and SO users (marginal travel time) are used that leads to a more even traffic distribution.

Under the spatially differentiated pricing scheme, UE-seeking HVs are more likely to take the second or third shortest path if the actual travel time on the first shortest path increases resulting in an NFD with a smaller hysteresis loop. NFDs presented in Fig. 12 confirm that the proposed pricing not only increases the experienced flow capacity of the pricing area, but also significantly reduces the size of the hysteresis loop.

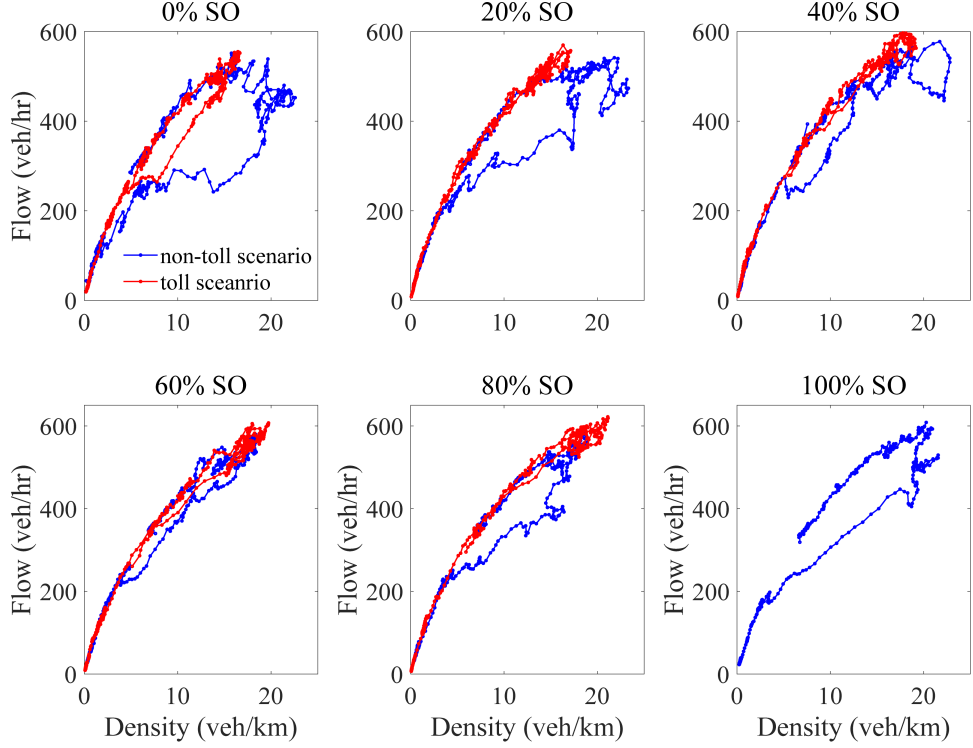

Fig. 12. NFD of the CBD in non-toll and toll scenarios: comparison of the size of the hysteresis loop representing heterogeneity of the spatial traffic distribution in the network.

Table 4 shows the average travel time of UE- and SO-seeking users in the CBD under routing only control for different scenarios. It also shows the average travel time of UE- and SO-seeking users under joint routing and pricing control as well as the average toll imposed on UE-seeking users and the benefit/cost ratio of UE users defined as the UE-seeking users travel time reduction due to the pricing divided by the average toll imposed on them to pass the pricing zone. Under routing only control, the cooperative SO users average travel time is only slightly higher than selfish UE users. However, this difference grows larger under joint routing and pricing control. The proposed spatially differentiated distance pricing not only reduces the inflow of UE users, but also drives them into the less congested links in the pricing zone. Thus, it brings lower average travel time for both UE and SO users in the pricing zone. The benefit/cost ratio suggests that it is a beneficial action to enter the pricing zone only when the SO ratio is almost less than 30% in the network.

Table 4. Comparison of travel time in the pricing zone under different control schemes for the different vehicle classes and SO ratios

| | Routing only control | | Joint routing and pricing control | | | | |
|---|---|---|---|---|---|---|---|
| SO ratio in the network (%) | UE users travel time (min) | SO users travel time (min) | UE users travel time (min) | SO users travel time (min) | Avg. toll imposed on UE users ($) | Avg. toll imposed on UE users (min*) | Benefit/cost ratio of UE users |
| 0 | 8.35 | - | 6.11 | - | 0.24 | 0.96 | 2.34 |
| 20 | 8.1 | 8.55 | 5.7 | 6.45 | 0.39 | 1.56 | 1.54 |
| 40 | 6.83 | 7.52 | 5.67 | 7.41 | 0.63 | 2.52 | 0.46 |
| 60 | 6.54 | 6.71 | 5.57 | 7.48 | 1.47 | 5.88 | 0.17 |
| 80 | 6.4 | 7.1 | 5.43 | 7.52 | 2.04 | 8.16 | 0.12 |
| 100 | - | 7.55 | - | 7.55 | 0 | 0 | - |

*The toll rate ($) is converted to (min) using VOT=15 $/hr.

Fig. 13 compares average network travel time (per unit distance) and average pricing zone density with routing control only scheme and joint routing and pricing control scheme. The travel time per unit distance measure is commonly used in performance comparison of large-scale network simulations (Chen et al., 2018, Jiang et al., 2011, Saedi et al., 2020). When the routing only control scheme is applied, the average travel time in the network decreases by increasing the SO ratio and the average density in the CBD remains roughly unchanged at 20 veh/km for different SO ratios. However, when the joint routing and pricing control scheme is applied and the proportion of SO-seeking vehicles in the network is relatively small, the pricing zone is less congested because a small number of vehicles can freely enter the pricing zone without paying toll while a large number of vehicles (UE-seeking users) are subject to a congestion charge and have the opportunity to detour and avoid entering the pricing zone. So, the applied feedback-based congestion pricing controller can set a reasonable toll and keep the pricing zone uncongested. In 0% SO scenario in which all users are subject to congestion charge for entering the pricing zone, the average pricing zone density reaches a minimum (15 veh/km) and in 20%, 40% and 60% SO scenarios keep fluctuating around 16-17 veh/km. By further increasing the SO ratio, traffic congestion in the pricing zone worsens and even imposing high toll rates cannot keep the pricing zone uncongested. The pricing zone density reaches 20 veh/km in 100% SO scenario where congestion pricing is practically terminated, and the toll would be zero. Like routing only control scheme, increasing SO ratio in joint routing and pricing control brings lower total travel time to the network. Therefore, there is a trade-off between reduction of the average travel time in the network and the average density in the pricing zone.

Results also suggest that the optimal ratio of SO-seeking users is between 40 and 60%, where the joint routing and pricing control scheme works effectively and meet both objectives to reasonably reduce total travel time in the network and congestion in the pricing zone. Exempting SO users from paying toll is shown to be an effective means to change road users' behavior. Increasing the SO ratio beyond 60% significantly increases the average pricing zone density and thus, impairs the effect of the pricing control. Therefore, proposing two different toll rates for different users need to be further investigated as an alternative pricing mechanism.

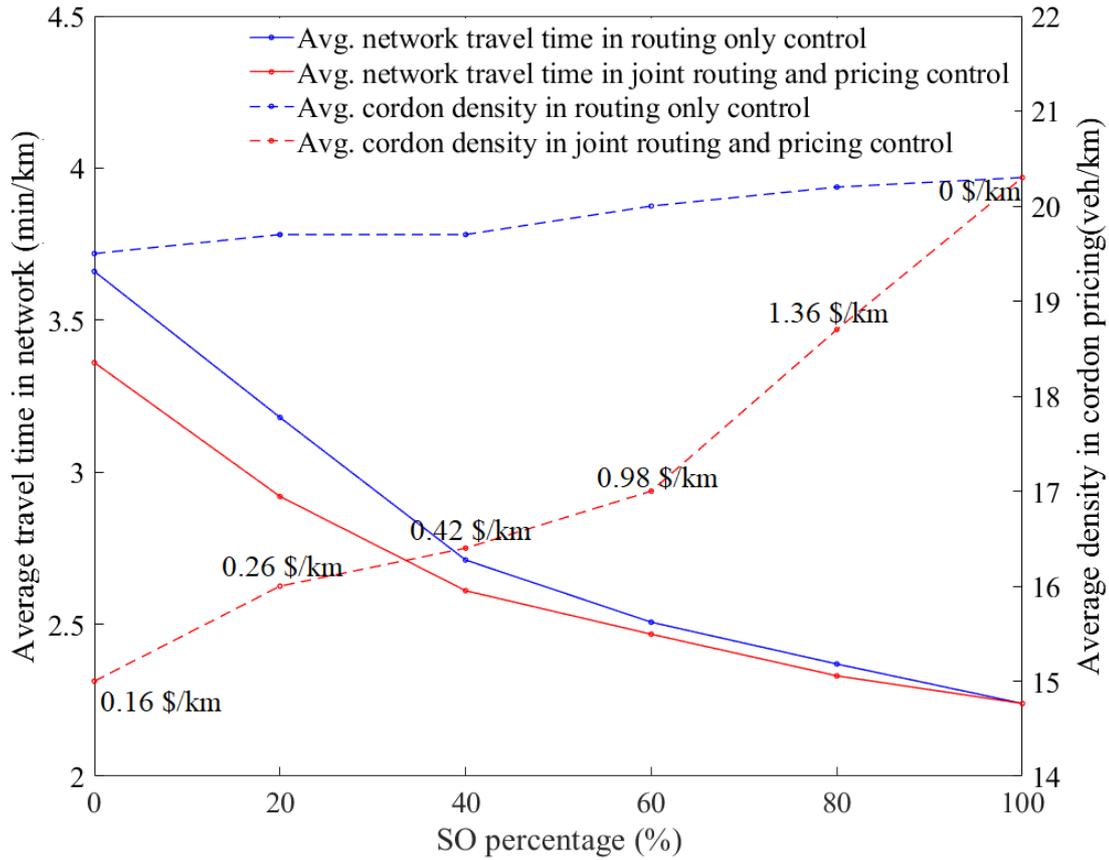

Fig. 13. Comparison of average network travel time per unit distance and average pricing zone density when routing only control is applied against joint routing and pricing control.

## 5. Conclusion

Despite rapid growth of the CAV technologies and their test implementations over the past few years, there could be a long transition period in which CAVs and HVs would share the transport infrastructure (i.e. mixed autonomy). Therefore, understanding and modeling the mixed traffic flow is necessary. As connectivity enables vehicle routing controllability (Rossi et al., 2017), SO-seeking CAVs are a conceivable mode as opposed to UE-seeking CAVs and HVs. However, it is questionable whether CAV owners or operators would be willing to be controlled externally by a central agency and follow SO routing without any incentive. Also, the potential increase in travel demand as a result of CAV ownership might exacerbate the traffic congestion in some parts of the network (Simoni et al., 2019, Levin and Boyles, 2016).

Unlike previous studies on congestion pricing of CAVs (Sharon et al., 2017, Simoni et al., 2019), this paper aimed to utilize and integrate a second-best and incentive-based congestion pricing scheme with a routing strategy. In the pricing strategy, SO-seeking CAVs are exempt from the tolls whereas UE-seeking users are subject to the charge for entering the pricing zone. This joint control scheme manages to encourage controllable CAVs to follow SO routing and discourage UE-seeking users from entering the congested city center.

The paper implements a mixed equilibrium simulation-based dynamic traffic assignment (SBDTA) framework for a mixed fleet of CAVs and HVs. A feedback-based congestion pricing controller is applied to keep the pricing zone away from congestion. The numerical results on the Melbourne network suggest that the total system travel time and the average network travel time decrease as the penetration rate of SO-seeking users in the network increases, as expected. For example, controlling 40% of CAVs (routing only control) leads to a 25% improvement in the average network travel time, although the city center is still congested (20 veh/km). The proposed joint routing and pricing control scheme not only reduces the average density in the CBD (16 veh/km), but also slightly improves the average network travel time because of the tolls. Results from both control schemes suggest that the size of the hysteresis loop in the NFD of the pricing zone is reduced significantly when the ratio of SO-seeking users is between 40% and 60%. It can reflect the homogenous distribution of the congestion in mixed traffic with mixed routing strategies as opposed to a single route strategy. Increasing the SO ratio beyond 60% may reduce the efficiency of the incentive-based pricing scheme, since the number of SO users in the pricing zone increases significantly and imposing high tolls on the remainder of UE users cannot keep the pricing zone uncongested.

In this study, the modal shares (SO-seeking CAVs and UE-seeking HVs) are fixed and non-uniformly distributed across the origins and destinations. The proposed spatially differentiated pricing scheme discourages UE-seeking HVs from travelling within the pricing zone, while SO-seeking CAVs are not subject to any charge throughout the network. Therefore, only the routes of UE-seeking HVs are affected by the incentive-based congestion pricing scheme and the SO- and UE-seeking ratios remain unchanged. Finding the optimal modal shares affected by the level of CAVs' compliance in response to the toll and traffic conditions requires a dynamic choice model as the third level in the proposed mathematical program. Solving a tri-level optimization problem is arguably too expensive, especially when dealing with dynamic large-scale networks and falls beyond the scope of the current paper. As a direction for future research, further investigation is needed to relax the inelastic demand assumption and extend the model to a multi-modal network where public transportation is present, similar to the approaches presented in Zheng et al. (2016), Simoni et al. (2019), Loder et al. (2019), Loder et al. (2017).

**Appendix A. Adaptive link fundamental diagram**

AIMSUN mesoscopic simulation uses a simplified Gipps car-following model. The simplification results in a link FD that estimates traffic flow characteristics at a mesoscopic level (see Tss (2014) for further details). The resulting triangular flow-density FD is as follows:

$$q_t(k) = min\left(V.k_t, \frac{1-Lk_t}{R}\right) = min\left(V.k_t, \frac{1}{R}(1-\frac{k}{k_{jam}})\right) \quad \text{A. 1}$$

$$R = R_{experiment}.R_{factor} \quad \text{A. 2}$$

The first term in the right-hand side of Equation A. 1 represents the undersaturated regime of the traffic where V and $k_t$ denote the speed limit and the density level of a given link. The second term represents the oversaturated regime where L and R is the effective length (sum of the vehicle length and the clearance distance) and the reaction time in a link, respectively. The link flow is zero either for k=0 or for $k_{jam}$=1/L and is maximum at the intersection of two regimes. Reaction time, here, is the production of the global reaction time and the reaction time factor of the link (Equation A. 2). The local reaction time factor brings more flexibility in the calibration process and is used to model links with specific situations including highly attractive environment, high slope or sharp bent that may affect the vehicle reaction time.

In this study, we use this factor in AIMSUN to produce adaptive link FDs and estimate the effects of a heterogeneous vehicle composition with different reaction times on link capacity and shockwave speed. As an example, Fig. A1 shows the adaptive FD of two links and the effect of shorter reaction time of CAVs on the link traffic flow. Fig. A1 (a) shows flow-density relationship of a single-lane link with jam density of 200 veh/km. As an example, the first observation is that the maximum experienced flow is 1200 veh/hr when all the link users are HVs. Also, a significant scatter appears around the oversaturated branch of the FD indicating that the link is highly congested in most of the simulation periods. While in 100% AV scenario, the maximum flow increases to 1800 veh/hr and the scatter in the congested regime reduces resulting in a more stable traffic flow. Fig. A1 (b) shows the FD of a four-lane highway link with jam density of 200 veh/km that follows the same trend.

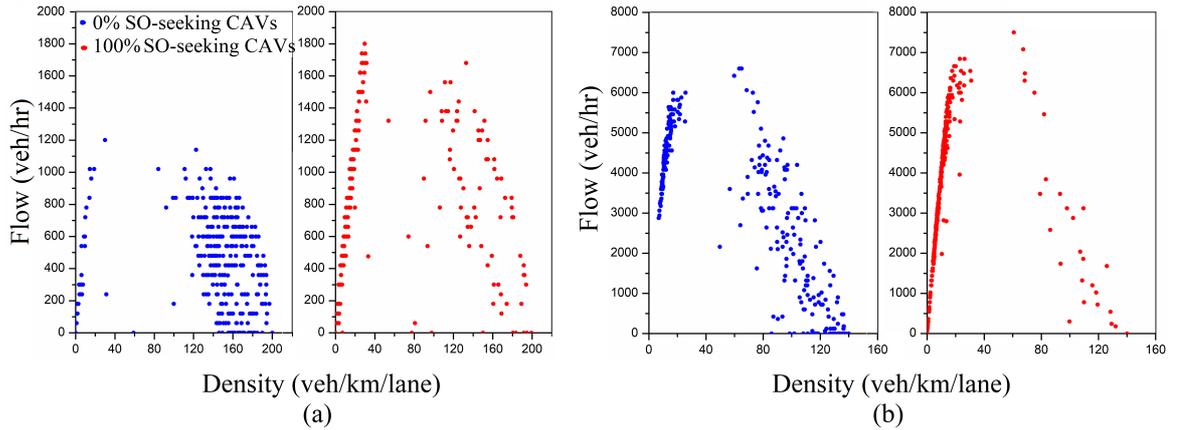

Fig. A1. Simulated flow-density diagram associated to full UE and SO scenarios for (a) a single lane section and (b) a 4-lane section.

**References**

Aboudina, A. & Abdulhai, B. 2017. A bi-level distributed approach for optimizing time-dependent congestion pricing in large networks: A simulation-based case study in the Greater Toronto Area. *Transportation Research Part C: Emerging Technologies,* 85**,** 684-710.

Aboudolas, K. & Geroliminis, N. 2013. Perimeter and boundary flow control in multi-reservoir heterogeneous networks. *Transportation Research Part B: Methodological,* 55**,** 265-281.

Bagloee, S. A., Sarvi, M., Patriksson, M. & Rajabifard, A. 2017. A mixed user‐equilibrium and system‐optimal traffic flow for connected vehicles stated as a complementarity problem. *Computer‐Aided Civil and Infrastructure Engineering,* 32**,** 562-580.

Ben-Akiva, M., De Palma, A. & Isam, K. 1991. Dynamic network models and driver information systems. *Transportation Research Part A: General,* 25**,** 251-266.

Bennett, L. D. 1993. The existence of equivalent mathematical programs for certain mixed equilibrium traffic assignment problems. *European Journal of Operational Research,* 71**,** 177-187.

Chen, X., Liu, Z., Zhang, K. & Wang, Z. 2020. A parallel computing approach to solve traffic assignment using path-based gradient projection algorithm. *Transportation Research Part C: Emerging Technologies,* 120**,** 102809.

Chen, X., Zhang, L., He, X., Xiong, C. & Li, Z. 2014. Surrogate‐based optimization of expensive‐to‐evaluate objective for optimal highway toll charges in transportation network. *Computer‐Aided Civil Infrastructure Engineering,* 29**,** 359-381.

Chen, X. M., Xiong, C., He, X., Zhu, Z. & Zhang, L. 2016. Time-of-day vehicle mileage fees for congestion mitigation and revenue generation: A simulation-based optimization method and its real-world application. *Transportation Research Part C: Emerging Technologies,* 63**,** 71-95.

Chen, Y.-J., Li, Z.-C. & Lam, W. H. 2018. Cordon toll pricing in a multi-modal linear monocentric city with stochastic auto travel time. *Transportmetrica A: Transport Science,* 14**,** 22-49.

Chen, Z., He, F., Yin, Y. & Du, Y. 2017. Optimal design of autonomous vehicle zones in transportation networks. *Transportation Research Part B: Methodological,* 99**,** 44-61.

Chiu, Y.-C., Bottom, J., Mahut, M., Paz, A., Balakrishna, R., Waller, T. & Hicks, J. 2011. Dynamic traffic assignment: A primer. *Transportation Research Circular*.

Daganzo, C. F. 1998. Queue spillovers in transportation networks with a route choice. *Transportation Science,* 32**,** 3-11.

Daganzo, C. F. & Lehe, L. J. 2015. Distance-dependent congestion pricing for downtown zones. *Transportation Research Part B: Methodological,* 75**,** 89-99.

Dong, C., Wang, H., Li, Y., Wang, W. & Zhang, Z. 2019. Route control strategies for autonomous vehicles exiting to off-ramps. *IEEE Transactions on Intelligent Transportation Systems,* 21**,** 3104-3116.

Ekström, J., Sumalee, A. & Lo, H. K. 2012. Optimizing toll locations and levels using a mixed integer linear approximation approach. *Transportation Research Part B: Methodological,* 46**,** 834-854.

Fagnant, D. J. & Kockelman, K. 2015. Preparing a nation for autonomous vehicles: opportunities, barriers and policy recommendations. *Transportation Research Part A: Policy and Practice,* 77**,** 167-181.

Fakhrmoosavi, F., Saedi, R., Zockaie, A. & Talebpour, A. 2020. Impacts of connected and autonomous vehicles on traffic flow with heterogeneous drivers spatially distributed over large-scale networks. *Transportation research record,* 2674**,** 817-830.

Fan, W. D. 2016. Optimal congestion pricing toll design under multiclass transportation network schemes: Genetic algorithm approaches. *Case Studies on Transport Policy,* 4**,** 78-87.

Florian, M. & Chen, Y. 1995. A Coordinate Descent Method for the Bi-level O–D Matrix Adjustment Problem. *International Transactions in Operational Research,* 2**,** 165-179.

Frank, M. & Wolfe, P. 1956. An algorithm for quadratic programming. *Naval research logistics quarterly,* 3**,** 95-110.

Friedrich, B. 2016. The effect of autonomous vehicles on traffic. *Autonomous Driving.* Springer.

Gayah, V. V. & Daganzo, C. F. 2011. Clockwise hysteresis loops in the macroscopic fundamental diagram: an effect of network instability. *Transportation Research Part B: Methodological,* 45**,** 643-655.

Geroliminis, N. & Daganzo, C. F. 2008. Existence of urban-scale macroscopic fundamental diagrams: Some experimental findings. *Transportation Research Part B: Methodological,* 42**,** 759-770.

Geroliminis, N., Haddad, J. & Ramezani, M. 2012. Optimal perimeter control for two urban regions with macroscopic fundamental diagrams: A model predictive approach. *IEEE Transactions on Intelligent Transportation Systems,* 14**,** 348-359.

Ghali, M. & Smith, M. 1995. A model for the dynamic system optimum traffic assignment problem. *Transportation Research Part B: Methodological,* 29**,** 155-170.

Gu, Z., Liu, Z., Cheng, Q. & Saberi, M. 2018a. Congestion pricing practices and public acceptance: A review of evidence. *Case Studies on Transport Policy,* 6**,** 94-101.

Gu, Z. & Saberi, M. 2021. Simulation-based optimization of toll pricing in large-scale urban networks using the network fundamental diagram: A cross-comparison of methods. *Transportation Research Part C: Emerging Technologies,* 122**,** 102894.

Gu, Z., Shafiei, S., Liu, Z. & Saberi, M. 2018b. Optimal distance- and time-dependent area-based pricing with the Network Fundamental Diagram. *Transportation Research Part C: Emerging Technologies,* 95**,** 1-28.

Gu, Z., Waller, S. T. & Saberi, M. 2019. Surrogate-based toll optimization in a large-scale heterogeneously congested network. *Computer - Aided Civil and Infrastructure Engineering,* 34**,** 638-653.

Haddad, J. 2017. Optimal perimeter control synthesis for two urban regions with aggregate boundary queue dynamics. *Transportation Research Part B: Methodological,* 96**,** 1-25.

Haddad, J., Ramezani, M. & Geroliminis, N. 2013. Cooperative traffic control of a mixed network with two urban regions and a freeway. *Transportation Research Part B: Methodological,* 54**,** 17-36.

Han, Y., Ramezani, M., Hegyi, A., Yuan, Y. & Hoogendoorn, S. 2020. Hierarchical ramp metering in freeways: an aggregated modeling and control approach. *Transportation research part C: emerging technologies,* 110**,** 1-19.

Han, Y., Wang, M., He, Z., Li, Z., Wang, H. & Liu, P. 2021. A linear Lagrangian model predictive controller of macro-and micro-variable speed limits to eliminate freeway jam waves. *Transportation Research Part C: Emerging Technologies,* 128**,** 103121.

Harker, P. T. 1988. Multiple equilibrium behaviors on networks. *Transportation science,* 22**,** 39-46.

Hu, T.-Y., Tong, C.-C., Liao, T.-Y. & Chen, L.-W. 2018. Dynamic route choice behaviour and simulation-based dynamic traffic assignment model for mixed traffic flows. *KSCE Journal of Civil Engineering,* 22**,** 813-822.

Huang, H.-J. & Li, Z.-C. 2007. A multiclass, multicriteria logit-based traffic equilibrium assignment model under ATIS. *European Journal of Operational Research,* 176**,** 1464-1477.

Janson, B. N. 1991. Dynamic traffic assignment for urban road networks. *Transportation Research Part B: Methodological,* 25**,** 143-161.

Jiang, L., Mahmassani, H. S. & Zhang, K. 2011. Congestion pricing, heterogeneous users, and travel time reliability: multicriterion dynamic user equilibrium model and efficient implementation for large-scale networks. *Transportation research record,* 2254**,** 58-67.

Keyvan-Ekbatani, M., Kouvelas, A., Papamichail, I. & Papageorgiou, M. 2012. Exploiting the fundamental diagram of urban networks for feedback-based gating. *Transportation Research Part B: Methodological,* 46**,** 1393-1403.

Le Vine, S., Liu, X., Zheng, F. & Polak, J. 2016. Automated cars: Queue discharge at signalized intersections with 'Assured-Clear-Distance-Ahead'driving strategies. *Transportation Research Part C: Emerging Technologies,* 62**,** 35-54.

Legaspi, J. & Douglas, N. Value of travel time revisited–NSW experiment.  Proc. 37th Australasian Transport Research Forum, 2015.

Levin, M. W. & Boyles, S. D. 2015. Effects of autonomous vehicle ownership on trip, mode, and route choice. *Transportation Research Record,* 2493**,** 29-38.

Levin, M. W. & Boyles, S. D. 2016. A multiclass cell transmission model for shared human and autonomous vehicle roads. *Transportation Research Part C: Emerging Technologies,* 62**,** 103-116.

Litman, T. 2017. *Autonomous vehicle implementation predictions*, Victoria Transport Policy Institute Victoria, Canada.

Liu, H. X., He, X. & He, B. 2009. Method of successive weighted averages (MSWA) and self-regulated averaging schemes for solving stochastic user equilibrium problem. *Networks Spatial Economics,* 9**,** 485.

Liu, Z., Meng, Q. & Wang, S. 2013. Speed-based toll design for cordon-based congestion pricing scheme. *Transportation Research Part C: Emerging Technologies,* 31**,** 83-98.

Liu, Z., Wang, S. & Meng, Q. 2014. Optimal joint distance and time toll for cordon-based congestion pricing. *Transportation Research Part B: Methodological,* 69**,** 81-97.

Loder, A., Ambühl, L., Menendez, M. & Axhausen, K. W. 2017. Empirics of multi-modal traffic networks–Using the 3D macroscopic fundamental diagram. *Transportation Research Part C: Emerging Technologies,* 82**,** 88-101.

Loder, A., Dakic, I., Bressan, L., Ambühl, L., Bliemer, M. C., Menendez, M. & Axhausen, K. W. 2019. Capturing network properties with a functional form for the multi-modal macroscopic fundamental diagram. *Transportation Research Part B: Methodological,* 129**,** 1-19.

Madadi, B., Van Nes, R., Snelder, M. & Van Arem, B. 2020. A bi‐level model to optimize road networks for a mixture of manual and automated driving: An evolutionary local search algorithm. *Computer‐Aided Civil Infrastructure Engineering,* 35**,** 80-96.

Mahmassani, H. S. 1994. Development and testing of dynamic traffic assignment and simulation procedures for ATIS/ATMS applications. Austin, Texas: Center for Transportation Research, The University of Texas at Austin.

Mahmassani, H. S. 2001. Dynamic network traffic assignment and simulation methodology for advanced system management applications. *Networks and spatial economics,* 1**,** 267-292.

Mahmassani, H. S. & Peeta, S. 1993. Network performance under system optimal and user equilibrium dynamic assignments: implications for advanced traveler information systems. *Transportation Research Record: Journal of the Transportation Research Board,* 1408**,** 83-93.

Mahmassani, H. S., Williams, J. C. & Herman, R. 1984. Investigation of network-level traffic flow relationships: some simulation results. *Transportation Research Record,* 971**,** 121-130.

Mohajerpoor, R., Saberi, M., Vu, H. L., Garoni, T. M. & Ramezani, M. 2020. H∞ robust perimeter flow control in urban networks with partial information feedback. *Transportation Research Part B: Methodological,* 137**,** 47-73.

Papageorgiou, M., Hadj-Salem, H. & Blosseville, J.-M. 1991. ALINEA: A local feedback control law for on-ramp metering. *Transportation Research Record: Journal of the Transportation Research Board,* 1320**,** 58-67.

Peeta, S. & Mahmassani, H. S. 1995. System optimal and user equilibrium time-dependent traffic assignment in congested networks. *Annals of Operations Research,* 60**,** 81-113.

Pigou, A. C. 1920. The economics of welfare. *Macmillan, London*.

Powell, W. B. & Sheffi, Y. 1982. The convergence of equilibrium algorithms with predetermined step sizes. *Transportation Science,* 16**,** 45-55.

Qian, Z. S., Shen, W. & Zhang, H. 2012. System-optimal dynamic traffic assignment with and without queue spillback: Its path-based formulation and solution via approximate path marginal cost. *Transportation research part B: methodological,* 46**,** 874-893.

Ramezani, M., Haddad, J. & Geroliminis, N. 2015. Dynamics of heterogeneity in urban networks: aggregated traffic modeling and hierarchical control. *Transportation Research Part B: Methodological,* 74**,** 1-19.

Ramezani, M. & Nourinejad, M. 2018. Dynamic modeling and control of taxi services in large-scale urban networks: A macroscopic approach. *Transportation Research Part C: Emerging Technologies,* 94**,** 203-219.

Rossi, F., Zhang, R., Hindy, Y. & Pavone, M. 2017. Routing autonomous vehicles in congested transportation networks: structural properties and coordination algorithms. *Autonomous Robots***,** 1-16.

Saberi, M., Mahmassani, H. S., Hou, T. & Zockaie, A. 2014. Estimating network fundamental diagram using three-dimensional vehicle trajectories: extending edie's definitions of traffic flow variables to networks. *Transportation Research Record,* 2422**,** 12-20.

Saedi, R., Saeedmanesh, M., Zockaie, A., Saberi, M., Geroliminis, N. & Mahmassani, H. S. 2020. Estimating network travel time reliability with network partitioning. *Transportation Research Part C: Emerging Technologies,* 112**,** 46-61.

Saffari, E., Yildirimoglu, M. & Hickman, M. 2020. A methodology for identifying critical links and estimating macroscopic fundamental diagram in large-scale urban networks. *Transportation Research Part C: Emerging Technologies,* 119**,** 102743.

Shafiei, S., Gu, Z. & Saberi, M. 2018. Calibration and validation of a simulation-based dynamic traffic assignment model for a large-scale congested network. *Simulation Modelling Practice and Theory,* 86**,** 169-186.

Sharon, G., Levin, M. W., Hanna, J. P., Rambha, T., Boyles, S. D. & Stone, P. 2017. Network-wide adaptive tolling for connected and automated vehicles. *Transportation Research Part C: Emerging Technologies,* 84**,** 142-157.

Sheffi, Y. 1985. Urban transportation networks: Equilibrium analysis with mathematical programming methods *(New Jersey: Prentice-Hall, Inc.)***,** pp. 55-60.

Sheffi, Y. & Powell, W. B. 1983. Optimal signal settings over transportation networks. *Journal of Transportation Engineering,* 109**,** 824-839.

Simoni, M., Pel, A., Waraich, R. A. & Hoogendoorn, S. 2015. Marginal cost congestion pricing based on the network fundamental diagram. *Transportation Research Part C: Emerging Technologies,* 56**,** 221-238.

Simoni, M. D., Kockelman, K. M., Gurumurthy, K. M. & Bischoff, J. 2019. Congestion pricing in a world of self-driving vehicles: An analysis of different strategies in alternative future scenarios. *Transportation Research Part C: Emerging Technologies,* 98**,** 167-185.
Sun, X., Liu, Z.-Y., Thompson, R. G., Bie, Y.-M., Weng, J.-X. & Chen, S.-Y. 2016. A multi-objective model for cordon-based congestion pricing schemes with nonlinear distance tolls. *Journal of Central South University,* 23**,** 1273-1282.
Tss 2014. Aimsun 8 Dynamic Simulators Users' Manual. *Barcelona, Spain.*
Van Vuren, T. & Watling, D. 1991. A multiple user class assignment model for route guidance. *Transportation Research Record***,** 22-22.
Verhoef, E. The economics of regulating road transport. *Books (1996)*.
Wadud, Z., Mackenzie, D. & Leiby, P. 2016. Help or hindrance? The travel, energy and carbon impacts of highly automated vehicles. *Transportation Research Part A: Policy and Practice,* 86**,** 1-18.
Wang, J., Peeta, S. & He, X. 2019. Multiclass traffic assignment model for mixed traffic flow of human-driven vehicles and connected and autonomous vehicles. *Transportation Research Part B: Methodological,* 126**,** 139-168.
Wang, Y., Szeto, W. Y., Han, K. & Friesz, T. L. 2018. Dynamic traffic assignment: A review of the methodological advances for environmentally sustainable road transportation applications. *Transportation Research Part B: Methodological,* 111**,** 370-394.
Wardrop, J. G. & Whitehead, J. I. 1952. Correspondence. some theoretical aspects of road traffic research. *Proceedings of the institution of civil engineers,* 1**,** 767-768.
Wu, W., Zhang, F., Liu, W. & Lodewijks, G. 2020. Modelling the traffic in a mixed network with autonomous-driving expressways and non-autonomous local streets. *Transportation Research Part E: Logistics Transportation Review,* 134**,** 101855.
Yang, H. 1995. Heuristic algorithms for the bilevel origin-destination matrix estimation problem. *Transportation Research Part B: Methodological,* 29**,** 231-242.
Yang, H. 1998. Multiple equilibrium behaviors and advanced traveler information systems with endogenous market penetration. *Transportation Research Part B: Methodological,* 32**,** 205-218.
Yang, H. & Huang, H.-J. 2004. The multi-class, multi-criteria traffic network equilibrium and systems optimum problem. *Transportation Research Part B: Methodological,* 38**,** 1-15.
Yang, H. & Huang, H.-J. 2005. *Mathematical and economic theory of road pricing*.
Yang, H. & Zhang, X. 2003. Optimal toll design in second-best link-based congestion pricing. *Transportation Research Record,* 1857**,** 85-92.
Yang, K., Zheng, N. & Menendez, M. 2017. Multi-scale perimeter control approach in a connected-vehicle environment. *Transportation research procedia,* 23**,** 101-120.
Zhang, K. & Nie, Y. M. 2018. Mitigating the impact of selfish routing: An optimal-ratio control scheme (ORCS) inspired by autonomous driving. *Transportation Research Part C: Emerging Technologies,* 87**,** 75-90.
Zhang, X. & Yang, H. 2004. The optimal cordon-based network congestion pricing problem. *Transportation Research Part B: Methodological,* 38**,** 517-537.
Zheng, N., Rerat, G. & Geroliminis, N. 2016. Time-dependent area-based pricing for multimodal systems with heterogeneous users in an agent-based environment. *Transportation Research Part C-Emerging Technologies,* 62**,** 133-148.
Zheng, N., Waraich, R. A., Axhausen, K. W. & Geroliminis, N. 2012. A dynamic cordon pricing scheme combining the Macroscopic Fundamental Diagram and an agent-based traffic model. *Transportation Research Part A: Policy and Practice,* 46**,** 1291-1303.
Ziliaskopoulos, A. K., Waller, S. T., Li, Y. & Byram, M. 2004. Large-scale dynamic traffic assignment: Implementation issues and computational analysis. *Journal of Transportation Engineering,* 130**,** 585-593.